\documentclass[twocolumn]{aastex62}

\usepackage{natbib}
\usepackage{graphicx}
\usepackage{float}
\usepackage{amsmath}
\usepackage[caption=false]{subfig}
\usepackage{booktabs}
\newcommand{\Ha}{H$\alpha$}
\newcommand{\Hb}{H$\beta$}
\newcommand{\bd}{H$\alpha$/H$\beta$}
\newcommand{\bdmath}{\mathrm{H}\alpha/\mathrm{H}\beta}
\newcommand{\mathHa}{\mathrm{H}\alpha}
\newcommand{\mathHb}{\mathrm{H}\beta}
\newcommand{\ebmvneb}{${E(B-V)_{\mathrm{neb}}}$}
\newcommand{\ebmvcont}{${E(B-V)_{\mathrm{cont}}}$}
\newcommand{\ebmvsed}{${E(B-V)_{\mathrm{SED}}}$}
\newcommand{\sfrsed}{$\mathrm{SFR}_{\mathrm{SED}}$}
\newcommand{\sfrha}{$\mathrm{SFR}_{\mathrm{H}\alpha}$}
\newcommand{\Zneb}{{12 + log(O/H)\textsubscript{O3N2}}}
\newcommand{\mstar}{$M_*$}
\defcitealias{reddy2016}{R16}
\defcitealias{bruzual2003}{BC03}

\shorttitle{Dust Attenuation, Star Formation, and Metallicity in \lowercase{$z \sim 2-3$} Galaxies}
\shortauthors{Theios et al.}

\begin{document}

\title{Dust Attenuation, Star Formation, and Metallicity in \lowercase{$z \sim 2-3$} Galaxies from KBSS-MOSFIRE\footnote{The data presented in this paper were obtained at the W.M. Keck Observatory, which is operated as a scientific partnership among the California Institute of Technology, the University of California, and the National Aeronautics and Space Administration.  The Observatory was made possible by the generous financial support of the W.M. Keck Foundation.}}
\author{Rachel L. Theios}
\affiliation{Cahill Center for Astronomy and Astrophysics, California Institute of Technology, 1200 East California Boulevard, MC 249-17 Pasadena, CA 91125, USA}

\author{Charles C. Steidel}
\affiliation{Cahill Center for Astronomy and Astrophysics, California Institute of Technology, 1200 East California Boulevard, MC 249-17 Pasadena, CA 91125, USA}

\author{Allison L. Strom}
\affiliation{Carnegie Observatories, 813 Santa Barbara Street, Pasadena, CA 91101, USA}

\author{Gwen C. Rudie}
\affiliation{Carnegie Observatories, 813 Santa Barbara Street, Pasadena, CA 91101, USA}

\author{Ryan F. Trainor}
\affiliation{Physics and Astronomy Department, Franklin \& Marshall College, 415 Harrisburg Pike, Lancaster, PA 17603}

\author{Naveen A. Reddy}
\affiliation{Department of Physics and Astronomy, University of California, Riverside, 900 University Avenue, Riverside, CA 92521, USA}

\email{rtheios@astro.caltech.edu}

\begin{abstract}
We present a detailed analysis of 317 $2.0 \leq z \leq 2.7$ star-forming galaxies from the Keck Baryonic Structure Survey (KBSS). Using complementary spectroscopic observations with Keck/LRIS and Keck/MOSFIRE, as well as spectral energy distribution (SED) fits to broadband photometry, we examine the joint rest-UV and rest-optical properties of the same galaxies, including stellar and nebular dust attenuation, metallicity, and star formation rate (SFR). The inferred parameters of the stellar population (reddening, age, SFR, and stellar mass) are strongly dependent on the details of the assumed stellar population model and the shape of the attenuation curve. Nebular reddening is generally larger than continuum reddening, but with large scatter. Compared to local galaxies, high-redshift galaxies have lower gas-phase metallicities (and/or higher nebular excitation) at fixed nebular reddening, and higher nebular reddening at fixed stellar mass, consistent with gas fractions that increase with redshift. We find that continuum reddening is correlated with 12 + log(O/H)\textsubscript{O3N2} at $3.0\sigma$ significance, whereas nebular reddening is correlated with only $1.1\sigma$ significance. This may reflect the dependence of both continuum reddening and O3N2 on the shape of the ionizing radiation field produced by the massive stars. Finally, we show that \Ha-based and SED-based estimates of SFR exhibit significant scatter relative to one another, and agree on average only for particular combinations of spectral synthesis models and attenuation curves. We find that the SMC extinction curve predicts consistent SFRs if we assume the sub-solar ($0.14Z_{\odot}$) binary star models that are favored for high-redshift galaxies.
\end{abstract}
\keywords{galaxies: evolution --- galaxies: high-redshift --- galaxies: star formation --- ISM: abundances --- ISM: dust, extinction --- ISM: H~\textsc{ii} regions}
\section{Introduction}\label{sec:intro}

It is well-established that much of the stellar mass in the universe formed at $z > 1$ and that star-forming systems dominate the galaxy population at this epoch \citep{madau2014}. Constraining details of the cosmic star formation rate evolution has long been a focus of extragalactic astrophysics, but it is a complex issue, as estimates of star formation rate (SFR) are subject to uncertainties in a wide variety of parameters, including dust attenuation corrections, the timescale of the current episode of star formation, the ionizing photon production rate (used to estimate SFRs from \Ha\ luminosities), and the assumed stellar initial mass function (IMF).

In order to recover the intrinsic SFR of a galaxy, one must first determine the effect of dust obscuration on the observations used to measure it. However, this determination is not always straightforward, as both the rest-UV stellar continuum and nebular emission lines can be used to estimate the SFR at high redshift (in cases where far-IR measurements are not available), and these two SFR indicators neccessitate different measures of the dust attenuation correction. Generally, such corrections involve measuring the dust reddening $E(B-V)$ relative to an SED template and translating it to an attenuation in magnitudes $A_{\lambda}$ assuming some attenuation curve, usually parameterized by the quantity $k(\lambda)$, which is defined such that $k(\lambda) = A_{\lambda}/E(B-V)$.

Attenuation affecting the far-UV stellar continuum has been estimated using a variety of methods, including the so-called IRX-$\beta$ relation, where the UV continuum slope $\beta$, sensitive to the reddening, is correlated with the ratio $L_{\mathrm{IR}}/L_{\mathrm{UV}}$ (IRX), which traces the attenuation \citep{calzetti1994,meurer1999,adelberger2000,reddy2006,reddy2010,buat2011,reddy2012a,reddy2012b}. Similarly, continuum attenuation can be measured by SFR comparisons \citep{erb2006,daddi2007,reddy2010,reddy2012b} and SED fitting to broad- and medium-band photometry using stellar population synthesis (SPS) models \citep{kriek2013,reddy2015}. At high redshift, the UV slope has been shown to be a reasonable tracer of dust attenuation on average \citep[e.g.][]{reddy2006,daddi2007,reddy2010,reddy2012a}, although it is also sensitive to the stellar population age, metallicity, star formation history, IMF, and binarity \citep[][and references therein]{reddy2015}. 

The standard method for measuring dust attenuation towards star-forming regions in local galaxies involves comparing the Balmer decrement (BD)---the observed ratio $I(\mathHa)/I(\mathHb)$---to the intrinsic value expected with no nebular reddening, $I(\mathHa)/I(\mathHb) = 2.86$ for $T_e = 10000$ K and Case B recombination \citep{osterbrock1989}. The Balmer decrement is used to derive a value of nebular reddening \ebmvneb, which is typically translated into a dust correction by assuming a line-of-sight extinction relation such as the Milky Way curve of \citet{cardelli1989}.
At $z > 1$, however, \Ha\ and \Hb\ are redshifted to near-IR wavelengths, where observations have historically been more difficult to obtain than in the optical. In recent years, with the advances made by slitless and multi-object near-IR spectrographs on the \emph{Hubble Space Telescope} (HST) and 8-10 m class ground-based telescopes, it has become feasible to measure the Balmer decrement directly for large samples of galaxies at intermediate and high redshift \citep[e.g.][]{price2014,reddy2015,nelson2016,strom2017}.

Although the general consensus is that the nebular reddening \ebmvneb\ should be used to derive a dust correction for emission lines and the continuum reddening \ebmvcont\ should be used to derive a dust correction for the continuum, the relationship between these two quantities remains uncertain at high redshift, as both values depend on the details of the assumed massive stellar populations. The detailed relationship between nebular and continuum reddening depends on complex effects such as geometry and the starburst age distribution.

There is an extensive body of literature comparing \ebmvneb\ and \ebmvcont\ \citep[e.g.][]{calzetti2000,kashino2013,price2014,reddy2015,shivaei2016}, with far from uniform results. \citet{calzetti2000} found that in nearby starburst galaxies, reddening is higher towards star-forming regions, and on average $E(B-V)_{\mathrm{neb}} = 2.27  E(B-V)_{\mathrm{cont}}$, where \ebmvcont\ is derived using the \citet{calzetti2000} starburst attenuation relation, and \ebmvneb\ assumes a line-of-sight relation such as the Galactic extinction curve of \citet{cardelli1989}\footnote{\citet{calzetti2000} used the \citet{seaton1979} extinction law.}. At intermediate and high redshift ($z \sim 1.5-3$), some studies have found that reddening is higher towards star-forming regions relative to the stellar populations producing the FUV-NUV continuum, where both the nebulae and the stellar continuum are assumed to be reddened using attenuation curves established at low redshift \citep[e.g.][]{kashino2013,price2014,reddy2015}. Other studies have found that the assumption $E(B-V)_{\mathrm{neb}} \approx E(B-V)_{\mathrm{cont}}$ is sufficient to produce consistent SFRs between observations of \Ha\ and X-ray, mid-IR, and far-IR observations of similarly selected galaxies at $z \sim 2$ \citep{reddy2004,erb2006,reddy2006,daddi2007,reddy2010,reddy2012a}. 

Many subsequent studies have used \ebmvcont\ as a proxy for \ebmvneb\ when direct measurements are not available. However, \citet{steidel2014} noted that the equation relating \ebmvcont\ to \ebmvneb\ given by \citet{calzetti2000} is often misinterpreted to mean that the attenuation in magnitudes affecting an \Ha\ photon is a factor of 2.27 higher than that affecting a continuum photon at the same wavelength. In fact, the equation $E(B-V)_{\mathrm{neb}} = 2.27 E(B-V)_{\mathrm{cont}}$ is valid only if the same attenuation curve (\citealt{calzetti2000}) and extinction curve (\citealt{cardelli1989}) have been used. Thus, the difference in attenuation between \Ha\ and the continuum depends on the details of the two assumed attenuation curves.

Another issue that remains to be addressed is the appropriate attenuation curve to apply to high-redshift galaxies. While some recent studies \citep[e.g.][]{alvarez-marquez2016,reddy2017} have found the SMC curve to best reproduce the conditions in high-redshift galaxies, other studies \citep[e.g.][]{mclure2018,koprowski2018} have argued that the \citet{calzetti2000} curve remains the most applicable attenuation curve.

Even after the appropriate dust correction has been determined, measurement of the intrinsic SFR of a galaxy depends on other parameters in a potentially complex manner. It is well-known that estimates of SFR for individual galaxies using different indicators can vary widely, particularly for the dustiest galaxies with $L_{\mathrm{bol}} \gtrsim 10^{12} L_{\odot}$ for which UV color-based methods systematically underpredict the SFR \citep[e.g.][]{reddy2012b}. Similarly, estimates of SFR may differ due to the timescale over which indicators are sensitive: \Ha\ luminosity is sensitive only to the stars producing significant numbers of ionizing photons, while the FUV continuum (1000-2000 \AA) may have significant contributions from stars with lifetimes up to $\sim 100$ Myr. In detail, converting an observation of \Ha\ luminosity or UV continuum slope to an SFR both depend on the timescale of the current episode of star formation and the star formation history (SFH) over the preceding $\sim 100$ Myr. 

Perhaps more importantly, estimates of the SFR depend on the nature of the population of massive stars that dominates the EUV and FUV light. Recent studies \citep[e.g.,][]{steidel2016} have indicated that developments in our understanding of massive stars, such as the significant role of binary evolution, may lead to revised estimates of galaxy SFRs at high redshift. For the BPASSv2.2 models \citep{stanway2018} used in this paper, the \Ha\ luminosity per unit solar mass of stars formed per year is larger by a factor of $\sim 2$ (once differences in IMF have been accounted for) compared to the canonical conversion often used in the literature \citep{kennicutt1998,kennicutt2012}. The canonical values are based on modeling of solar-metallicity nebulae (and single-star SPS models) that are likely appropriate for most $z \sim 0$ galaxies; however, the ISM conditions at $z \sim 2$ are quite different, characterized by lower gas-phase metallicities than $z\sim0$ galaxies at fixed stellar mass and harder ionizing radiation fields \citep{strom2017}. Thus, comparing SFRs estimated from \Ha\ luminosity, dust-corrected using the Balmer decrement, to those estimated from the UV stellar continuum, dust-corrected using the UV continuum slope, can yield important insights into the nature of star formation at high redshift.  

In this paper we analyze a sample of 317 galaxies with $2.0 \leq z \leq 2.7$ observed as part of the Keck Baryonic Structure Survey \citep[KBSS;][]{rudie2012,steidel2014,strom2017}, with high-quality near-IR spectra from Keck/MOSFIRE \citep{mclean2012} together with deep optical spectra obtained with Keck/LRIS \citep{oke1995,steidel2004}, providing complementary rest-frame optical and rest-frame UV spectra of the same objects. KBSS is a large, targeted spectroscopic survey designed to jointly probe galaxies and the surrounding intergalactic medium (IGM) and circumgalactic medium (CGM) at the peak epoch of galaxy assembly, $z \sim 2-3$. The redshift range $2.0 \leq z \leq 2.7$ is ideal for ground-based observations due to the fortuitous placement of nebular emission lines with respect to atmospheric transmission windows. Additionally, at $2.0 \leq z \leq 2.7$, the FUV portion of galaxy spectra falls above the atmospheric cutoff near 3100 \AA. Thus, this redshift range uniquely provides access to both the stellar FUV continuum and nebular emission lines, making it possible to directly compare stellar and nebular measures of reddening and star formation. Finally, the availability of strong nebular emission lines in this redshift range allows us to estimate gas-phase oxygen abundance using strong-line methods, as a function of stellar mass ($M_*$), SFR, and dust reddening.

The outline of the paper is as follows. In Section \ref{sec:obs} we introduce the subset of the KBSS-MOSFIRE sample discussed in this paper, and in Section \ref{sec:params} we describe our methodology for modeling stellar populations and measuring galaxy parameters. In Section \ref{sec:rest-uv} we present composite rest-UV spectra to which we have fit spectral synthesis models. We examine the relationship between stellar and nebular dust attenuation in Section \ref{sec:ebmv_comp}. We compare dust attenuation with inferred gas-phase oxygen abundance and stellar mass in Section \ref{sec:o3n2_bd}. Section \ref{sec:sfr_comp} compares SFRs inferred from \Ha\ with those inferred from SED fitting. Section \ref{sec:disc} discusses the implications of these results, and Section \ref{sec:summary} summarizes our conclusions. Appendix \ref{sec:sed} describes details of our SED fitting methodology, and Appendix \ref{sec:fitting} describes our methodology for fitting spectral synthesis models to rest-UV spectra.

Throughout the paper, we adopt a $\Lambda$CDM cosmology with $H_0 = 70~\mathrm{km}~\mathrm{s}^{-1}~\mathrm{Mpc}^{-1}$, $\Omega_{\Lambda} = 0.7$, and $\Omega_{\mathrm{m}} = 0.3$. Conversion relative to solar metallicity assumes $Z_{\odot}= 0.0142$ (where $Z$ is the fraction of metals by mass), as in \citet{asplund2009}. Specific spectral features are referred to using their vacuum wavelengths, and magnitudes are given in the AB system. 

\section{Observations and Data}\label{sec:obs}

\subsection{Photometric observations}\label{sec:phot}

Photometric data in the KBSS survey fields include broadband photometry in the optical ($U_nG\mathcal{R}$), near-IR ($J, H, K_s$, and WFC3-IR F140W and F160W), and mid-IR (\emph{Spitzer}-Infrared Array Camera $3.6\micron$ and $4.5\micron$). The near-IR photometry was corrected for the emission line contribution from \Ha\ and [O~\textsc{iii}] using measured fluxes from MOSFIRE. Most KBSS galaxies are selected by their rest-UV colors using a $U_nG\mathcal{R}$ color selection scheme designed to select Lyman Break Galaxy analogues at $z \gtrsim 2$. While this color selection technique may bias the sample against, massive, dusty galaxies, the inclusion of $\mathcal{R}K$ objects (discussed by \citealt{strom2017}) alleviates this bias somewhat. Details of the photometric data and sample selection are described elsewhere \citep[e.g.][]{adelberger2004,steidel2004,reddy2012b,steidel2014,strom2017}.

\subsection{LRIS observations}\label{sec:lris}
Optical (rest-UV) spectra of KBSS galaxies were obtained with Keck/LRIS. LRIS observations in the KBSS survey fields have been obtained over the course of the past decade, and include observations with a variety of wavelength coverage, signal to noise ratio, and spectral resolution \citep[for details, see][]{shapley2005,steidel2010,reddy2012b,steidel2016}. Here we used observations with the blue channel of LRIS (LRIS-B), using both the 600 line/mm grism, which achieves a resolving power of $R \sim 1400$ and covers the wavelength range $3300-5600$ \AA, and the 400 line/mm grism, which achieves $R \sim 800$, covers $3100-6200$ \AA, and is optimized for the highest throughput at wavelengths $\leq 4000$ \AA.

Due to the variable signal to noise ratio of the LRIS spectra used in this paper, we stack the spectra in bins of selected galaxy properties (see Appendix \ref{sec:fitting} for details). For the stacking analysis, the spectra were interpolated onto a common rest wavelength scale of 0.35~\AA/pixel and averaged without weighting; the rest wavelength range $1100 \le \lambda_0 \le 1700$ \AA\ was used for fitting. To test that the relative flux calibration of the LRIS spectra was consistent with the photometry, we generated composite spectra in bins of $E(B-V)$ (as inferred from SED fits to broadband photometry of individual objects), and fit stellar spectral synthesis models to the composites to infer a continuum $E(B-V)$. The two measures of continuum reddening are comparable.

\subsection{MOSFIRE observations}\label{sec:mosfire}
Near-IR (rest-optical) spectra of KBSS galaxies were obtained with Keck/MOSFIRE. The near-IR portion of the KBSS survey (KBSS-MOSFIRE) was designed to provide high-quality rest-optical spectra for galaxies in the KBSS survey fields with multiwavelength ancillary data as well as spectroscopic observations (primarily with LRIS-B). For a full description of KBSS-MOSFIRE observing strategies, data reduction, sample selection, and emission-line fitting, see \citet{steidel2014} and \citet{strom2017}. In this work, we focus on the subset of KBSS-MOSFIRE galaxies with nebular redshifts $2.0 \lesssim z \lesssim 2.7$ in order to place important strong emission lines in the NIR atmospheric passbands, and with sufficiently deep \emph{H}-band and \emph{K}-band spectra to allow significant detections of the emission lines of interest: \Ha\ $\lambda6564.61$ and [N~\textsc{ii}]$\lambda6585.27$ in the \emph{K} band, and \Hb\ $\lambda4862.72$ and [O~\textsc{iii}]$\lambda5008.24$ in the \emph{H} band. The Balmer lines have been corrected for underlying Balmer absorption based on the best-fit SED models, as detailed by \citet{strom2017}.

Of particular importance for the purposes of this paper is the issue of cross-band calibration---correcting emission line fluxes for relative slit losses so that the ratios of lines observed in different spectral bands are accurate. This cross-calibration is necessary for analyses involving the Balmer decrement, as \Hb\ falls in $H$ band and \Ha\ in $K$ band for galaxies over the redshift range of interest for this study, $2.0 \leq z \leq 2.7$. The method used to estimate slit losses is described in detail by \citet{strom2017}. In brief, the cross-calibration combines separate observations of the same objects on independent slit masks with observations of a bright star placed on each mask in order to correct for slit losses for each object in each observed band.
	
\subsection{Sample statistics}\label{sec:sample}

\begin{deluxetable}{lDDD}
\tablecaption{Sample Statistics\label{tab:sample}}
\tablehead{\colhead{Galaxy Sample} & \multicolumn2c{$N_{\mathrm{gal}}$} & \multicolumn2c{$z_{\mathrm{neb}}$} & \multicolumn2c{$\bdmath$} \\ 
\colhead{} & \multicolumn2c{} & \multicolumn2c{(Mean)} & \multicolumn2c{(Median)}} 
\decimals
\startdata
KBSS-MOSFIRE                                       & 1103 & 2.23 &  . \nodata \\
$2.0 \leq z \leq 2.7$                              &  735 & 2.30 &  . \nodata \\
\Ha+\Hb\tablenotemark{a}                           &  373 & 2.29 & 3.97\pm1.90 \\
$\bdmath > 3\sigma$\tablenotemark{b}               &  317 & 2.29 & 3.87\pm1.60 \\\relax
[O~\textsc{iii}]+[N~\textsc{ii}]\tablenotemark{c}  &  270 & 2.29 & 3.88\pm1.60 \\
LRIS+MOSFIRE\tablenotemark{d}                      &  270 & 2.30 & 3.95\pm1.83 \\
\enddata
\tablenotetext{a}{The subset of KBSS-MOSFIRE galaxies with $2.0 \leq z \leq 2.7$ and measurements of the Balmer decrement \bd\ at any significance.}
\tablenotetext{b}{The main sample discussed in this paper is the subset of KBSS-MOSFIRE galaxies with $2.0 \leq z \leq 2.7$ and $S/N > 3$ on the Balmer decrement \bd, including slit loss uncertainties in \emph{H} and \emph{K}.}
\tablenotetext{c}{The [O~\textsc{iii}]+[N~\textsc{ii}] sample is the subset of those galaxies with observations of [O~\textsc{iii}]$\lambda5008$ and [N~\textsc{ii}]$\lambda6585$, 32 of which are undetected ($< 2\sigma$) in [N~\textsc{ii}]$\lambda6585$.}
\tablenotetext{d}{The LRIS+MOSFIRE sample is the subset of galaxies with complementary LRIS observations.}
\tablecomments{Error bars on median values are the inter-quartile range.}
\end{deluxetable}

\begin{figure}[tb]
\centering
\plotone{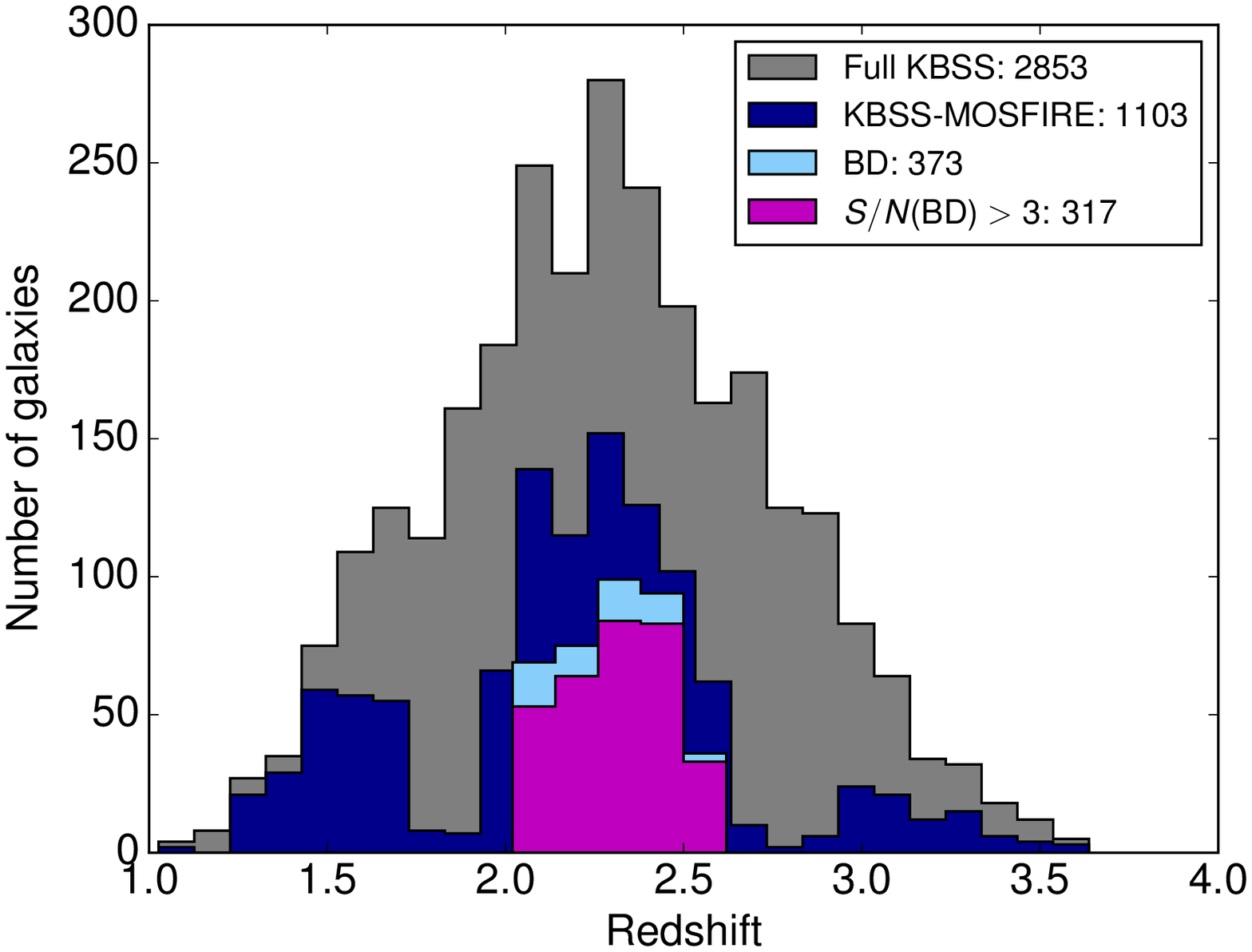}
\caption{Histogram of the galaxy samples discussed in this paper (see Table \ref{tab:sample}). \label{fig:zhist}}
\end{figure}
	
The full KBSS sample consists of 2844 galaxies with spectroscopically confirmed redshifts in the range $1.5 \la z \la 3.5$, 2345 of which have existing rest-frame UV spectra from Keck/LRIS-B. As of 2016 December 1, the near-IR portion of KBSS (KBSS-MOSFIRE) consists of 1103 galaxies with nebular redshifts obtained from Keck/MOSFIRE. AGN have been removed from the sample, as discussed by \citet{steidel2014} and \citet{strom2017}.

Table \ref{tab:sample} and Figure \ref{fig:zhist} give an overview of the galaxy samples discussed in this paper. We selected 317 KBSS-MOSFIRE galaxies in the redshift range $2.0 \leq z \leq 2.7$ with robust detections of both \Ha\ and \Hb\ ($S/N > 3$ for the line ratio \bd). The motivation for a significance cut on the Balmer decrement is discussed by \citet{strom2017}: since attenuation corrections scale nonlinearly with the measured value of the Balmer decrement, relatively small uncertainties in the Balmer decrement can translate to a large uncertainty in quantities that are dust-corrected using the Balmer decrement, such as the SFR. In practice, requiring  $S/N > 3$ in the ratio $\mathrm{BD} \equiv I(\mathHa)/I(\mathHb)$, including the relative slit loss uncertainties in \emph{H} and \emph{K}, means that $S/N > 3$ in \Ha\ and \Hb\ individually. While it is possible that this significance cut biases the sample against the most heavily reddened objects, the median BD of the $3\sigma$ sample and the subset of 373 galaxies with measurements of BD at any significance are consistent within the errors (Table \ref{tab:sample}).

\section{Derived galaxy properties}\label{sec:params}

\subsection{SED fitting}

Stellar masses, SFRs (hereafter \sfrsed), and continuum color excesses (hereafter \ebmvsed) were estimated for KBSS-MOSFIRE galaxies based on SED fits to broadband photometry described in Section \ref{sec:phot}.

The SED fitting uses reddened ``Binary Population and Spectral Synthesis'' \citep[BPASSv2.2;][]{stanway2018} models assuming a constant star formation history (SFH) and a minimum allowed age of 50 Myr. As discussed by \citet{reddy2012a}, this minimum age is approximately the central dynamical timescale and is imposed to prevent best-fit solutions with unrealistically young ages \footnote{While galaxies at these redshifts would likely have a rising SFH in reality, we expect the results assuming constant SFHs to be similar \citep{reddy2012a}. Rising SFHs will be implemented in a future paper.}. The BPASSv2.2 model used as the fiducial model for the SED fitting has a stellar metallicity of $Z_* = 0.002$ ($Z_*/Z_{\odot}\approx0.14$) and binary evolution included. The model we used has a Kroupa-type IMF, with a slope of $-2.35$ for masses $> 1 M_{\odot}$ and an upper mass cutoff of $100 M_{\odot}$. 

The main feature of the BPASSv2.2 models is the inclusion of massive interacting binary stars, increasing the overall ionizing flux and producing a harder ionizing spectrum for models with continuous star formation histories. This is an advantage over previous models not only because most massive stars are known to be in binaries \citep{crowther2007,sana2012} but because previous studies \citep[e.g.][]{steidel2016,reddy2016b,strom2017} have found that these models are better able to simultaneously match the rest-UV stellar continuum and nebular emission lines than single-star models. Additionally, \citet{steidel2016} noted that an important difference between the continuous star formation BPASS models including binary evolution and single-star models such as Starburst99 \citep{leitherer2014} and the single-star BPASS models is that the binary models predict a broad stellar He~\textsc{ii} $\lambda1640$ emission line (for constant SF models) not present in single-star models. This feature is detected in all our rest-UV composite spectra; see Section \ref{sec:rest-uv}. We elect to employ the BPASSv2.2 models over the more commonly used \citet[][hereafter BC03]{bruzual2003} solar metallicity models for these reasons.

The SED fitting procedure, characteristic uncertainties for each SED fit parameter, and comparisons to other stellar population synthesis (SPS) models are discussed in Appendix \ref{sec:sed}.

\subsection{Attenuation curves}\label{sec:curves}

\begin{figure}[tb]
	\centering
	\plotone{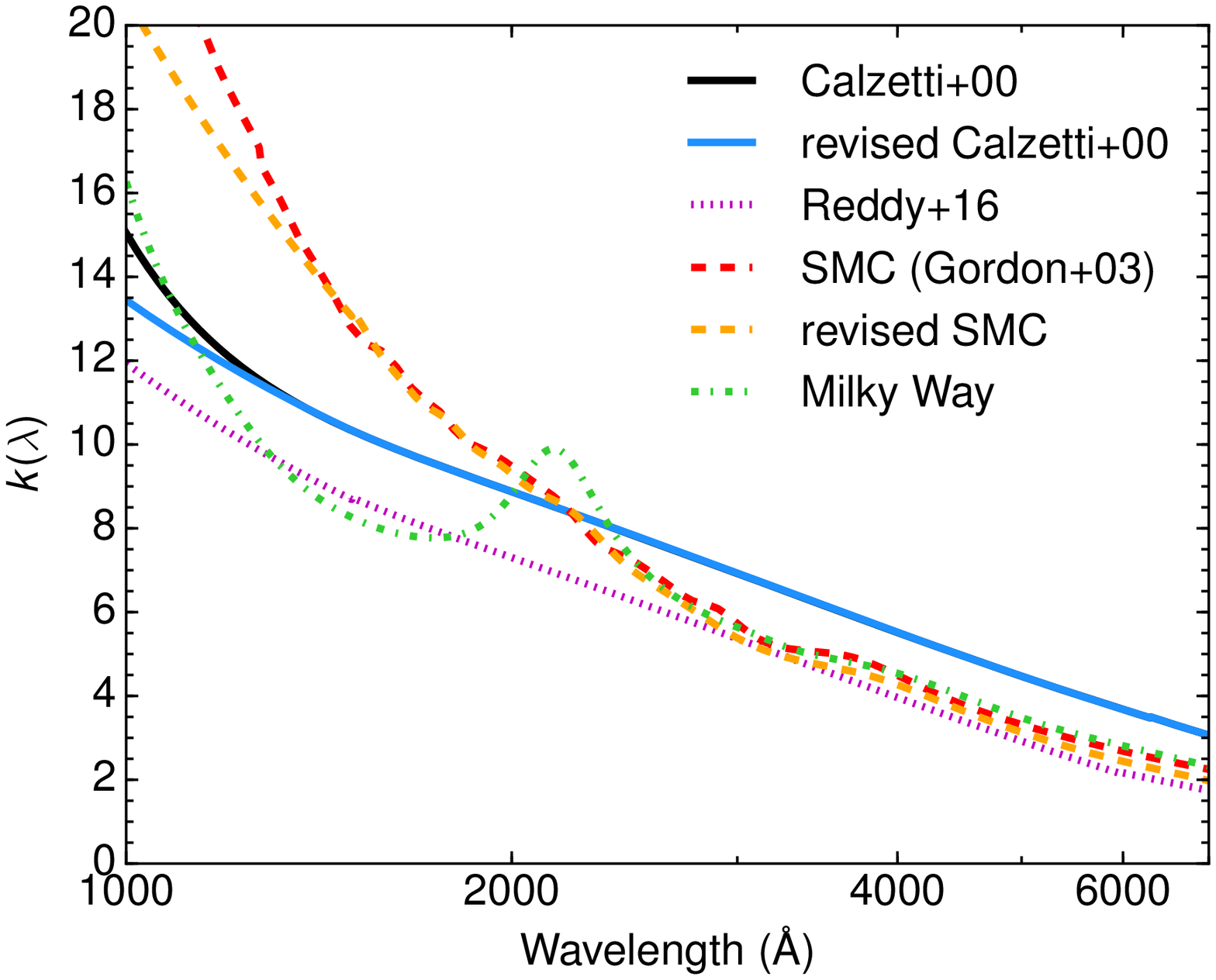}
	\caption{Comparison of attenuation curves discussed in this paper: the \citet{calzetti2000} starburst attenuation relation, the attenuation curve of \citet{reddy2016}, the SMC \citep{gordon2003}, and the Milky Way extinction curve of \citet{cardelli1989}. The revised \citet{calzetti2000} and SMC curves, used in this paper, include an empirical extension to the far-UV, using the method described in \citet{reddy2016}.\label{fig:klambda}}
	\end{figure}

We fit SEDs to each galaxy assuming two attenuation curves for the stellar continuum: \citet{calzetti2000} and SMC. For the nebular emission lines, we assume the Milky Way curve of \citet[][with $R_V = 3.1$]{cardelli1989}, or SMC for comparison when the continuum attenuation curve is assumed to be SMC.\footnote{While the choice of attenuation curve used to correct the stellar continuum has a large effect on the SED-inferred SFRs, it makes very little difference which curve is used to correct nebular emission line ratios, as all the commonly employed curves have a very similar slope and normalization in the optical regime (see Figure \ref{fig:klambda}). Note however that the due to the difference in $R_{\mathrm{V}}$}. The ``SMC'' curve used here combines the empirical SMC extinction curve from \citet[][assuming $R_V = 2.74$]{gordon2003} with an empirical extension to the far-UV using the method described by \citet{reddy2016}. The version of the \citet{calzetti2000} curve we use here has also been revised to include an extension to the far-UV using the same method. Similarly, we fit SEDs using an attenuation curve that combines the results of \citet{reddy2015} based on the MOSDEF survey \citep{kriek2015} with a new empirical extension to the far-UV described by \citet{reddy2016}. The results using this curve are similar to \citet{calzetti2000}, so for simplicity we discuss only \citet{calzetti2000} and SMC in this paper. The four curves discussed here are compared in Figure \ref{fig:klambda}. The original versions of the \citet{calzetti2000} and SMC curves are shown for reference.

\begin{figure}[htb]
\centering
\plotone{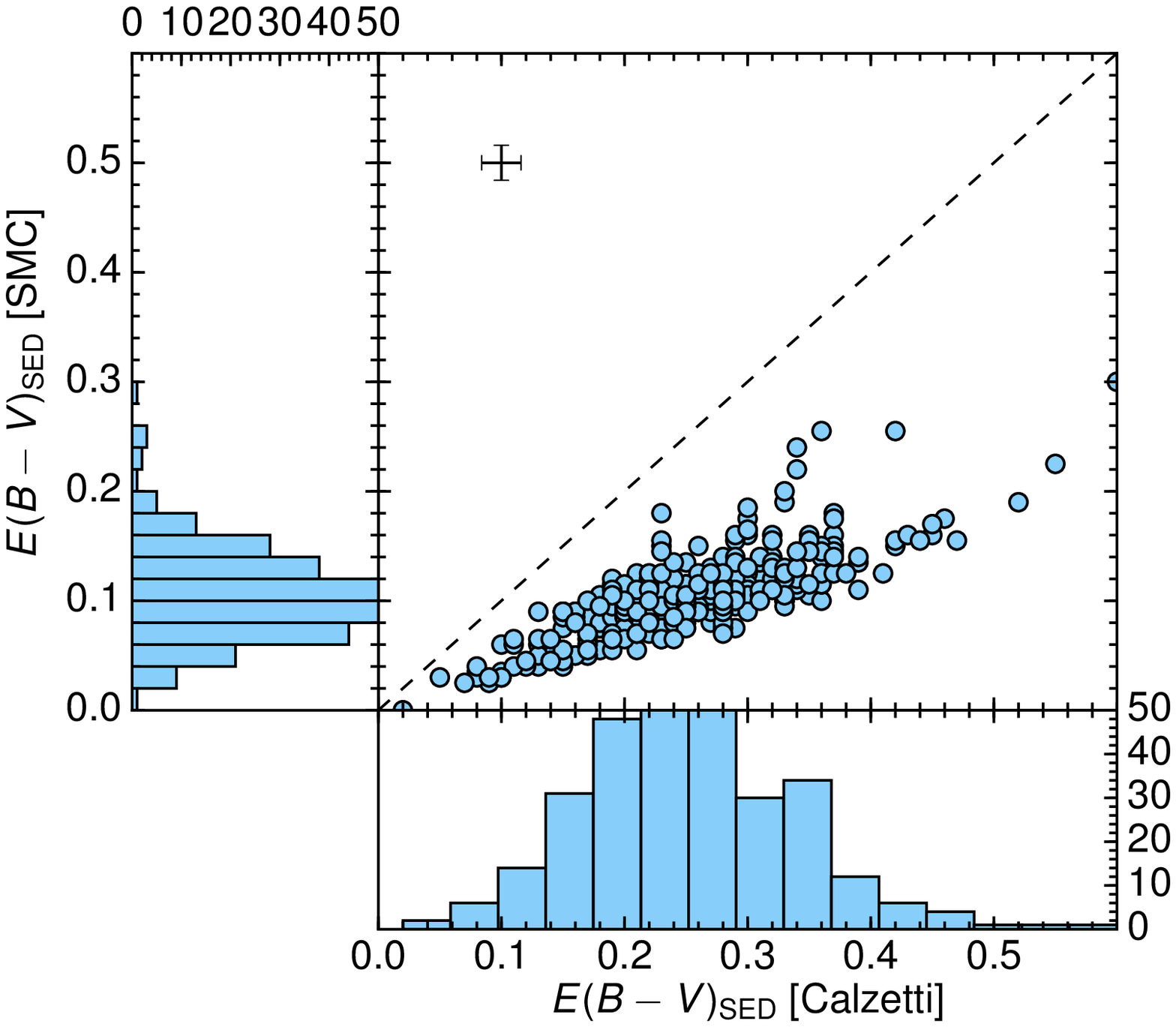}
\caption{\ebmvsed\ assuming the \citet{calzetti2000} attenuation curve versus \ebmvsed\ assuming the SMC extinction curve, for the 317 galaxies with $> 3\sigma$ measurements of BD (including the uncertainties on the slit loss corrections in $H$ and $K$). Here all quantities are derived from a BPASSv2.2 $Z_* = 0.002$ binary model. SMC predicts a lower \ebmvsed\ by a factor of 2.4 on average.\label{fig:ebmvsed}}
\end{figure}

\begin{figure}[htb]
\centering
\plotone{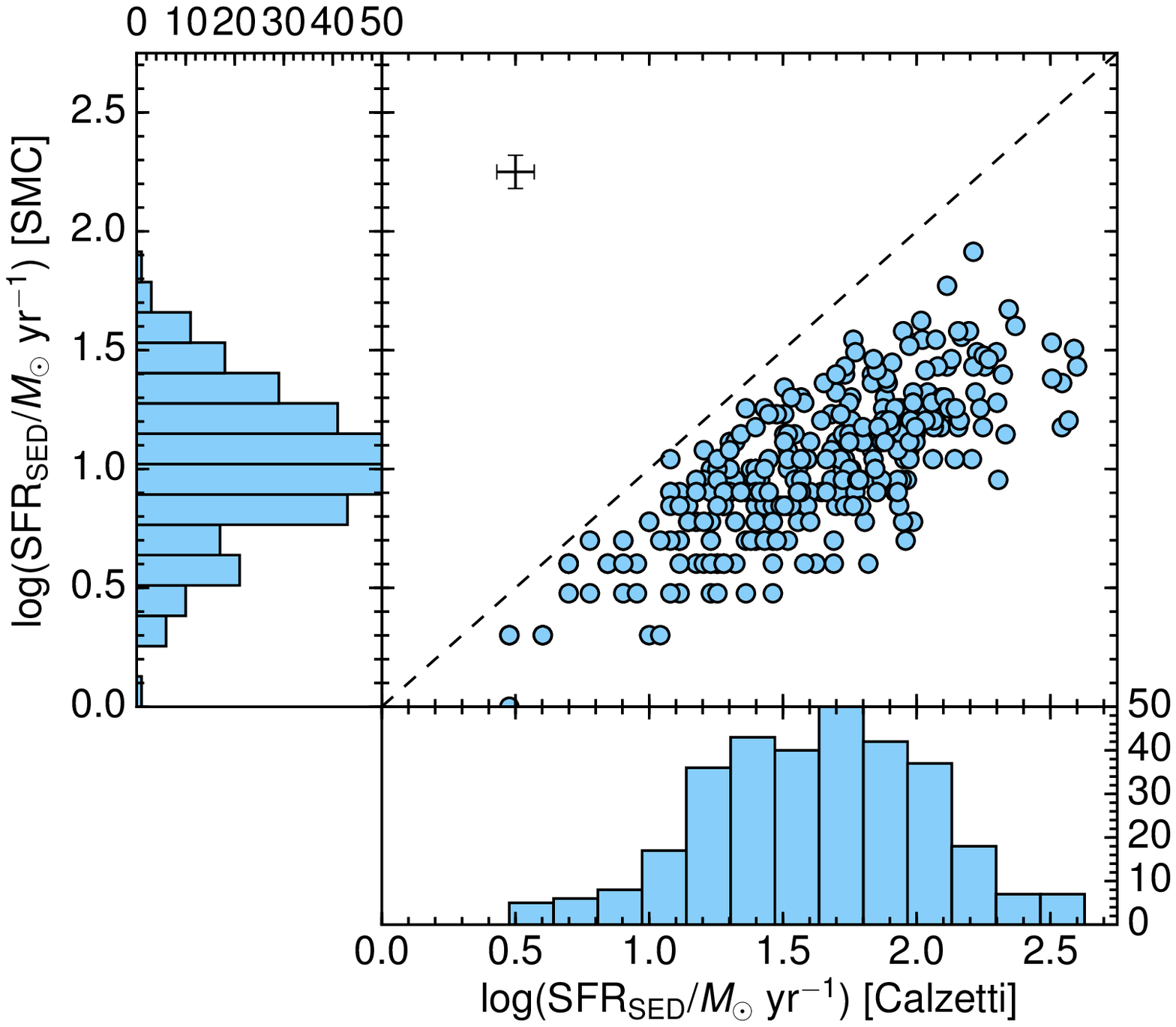}
\caption{\sfrsed\ assuming the \citet{calzetti2000} attenuation curve versus \sfrsed\ assuming the SMC extinction curve, for the 317 galaxies with $> 3\sigma$ measurements of BD (including the uncertainties on the slit loss corrections in $H$ and $K$). Here all quantities are derived from a BPASSv2.2 $Z_* = 0.002$ binary model. SMC predicts a lower \sfrsed\ by 0.50 dex on average.\label{fig:sfrsed}}
\end{figure}

\begin{figure}[htb]
\centering
\plotone{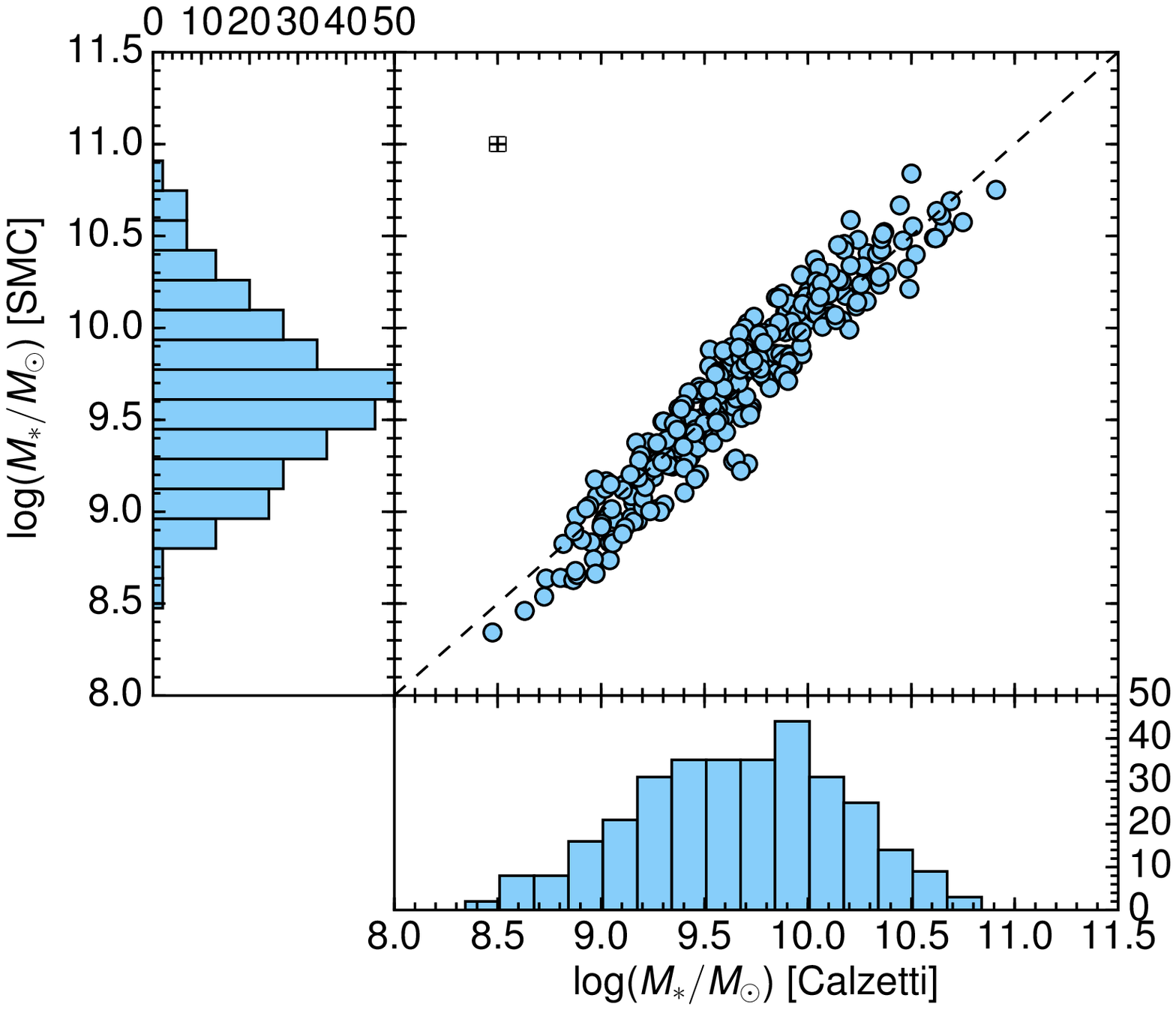}
\caption{\mstar\ assuming the \citet{calzetti2000} attenuation curve versus \mstar\ assuming the SMC extinction curve, for the 317 galaxies with $> 3\sigma$ measurements of BD (including the uncertainties on the slit loss corrections in $H$ and $K$). Here all quantities are derived from a BPASSv2.2 $Z_* = 0.002$ binary model. The stellar masses predicted by both curves are consistent on average.\label{fig:mstar}}
\end{figure}

Figures \ref{fig:ebmvsed}, \ref{fig:sfrsed}, and \ref{fig:mstar} compare the distributions of \ebmvsed, \sfrsed, and $M_*$, for \citet{calzetti2000} and SMC attenuation. While the inferred stellar masses are consistent within the errors, using SMC results in a lower \ebmvsed\ and \sfrsed\ due to the greater degree of attenuation per unit reddening in the UV. On average, \ebmvsed\ inferred for SMC is lower than that inferred for \citet{calzetti2000} by a factor of 2.4, and \sfrsed\ is lower by a factor of 3.3, or 0.5 dex. 

We find that 38\% of galaxies in the sample are better fit by SMC, 16\% are better fit by \citet{calzetti2000}, and 46\% are comparably fit by either curve within the uncertainty. Details and example SED fits are given in Appendix \ref{sec:sed-fit}. Throughout the paper, we compare results assuming each of the two curves.

\begin{deluxetable*}{lDDD}
\tablecaption{Star Formation Rate Calibrations\label{tab:sfr}}
\tablehead{\colhead{Model} & \multicolumn2c{$\log(L_{\mathrm{H}\alpha}/M_{\odot}~\mathrm{yr}^{-1})$} & \multicolumn2c{$\log(\nu L_{\nu}/M_{\odot}~\mathrm{yr}^{-1})$ [1500 \AA]} & \multicolumn2c{$\xi_{\mathrm{ion}}$}}
\decimals
\startdata
BPASSv2.2 $300M_{\odot}$ $Z_*=0.002$ & 41.78 & 43.51 & 25.44 \\
BPASSv2.2 $100M_{\odot}$ $Z_*=0.002$ & 41.64 & 43.46 & 25.35 \\
BPASSv2.2 $100M_{\odot}$ $Z_*=0.004$ & 41.59 & 43.46 & 25.30 \\
BPASSv2.2 $100M_{\odot}$ $Z_*=0.020$ & 41.35 & 43.36 & 25.16 \\
\citetalias{bruzual2003} $Z_*=0.004$ & 41.37 & 43.44 & 25.10 \\
\citetalias{bruzual2003} $Z_*=0.020$ & 41.14 & 43.40 & 24.91 \\
\citet{kennicutt2012}                & 41.30 & 43.35 & 25.12 \\
\enddata
\tablecomments{All quantities are corrected to the default BPASS IMF. BPASSv2.2 and \citetalias{bruzual2003} values assume a star formation timescale of $t = 10^{8.0}$ yr.}
\end{deluxetable*}

\subsection{Nebular reddening and \Ha\ SFR}\label{sec:neb}

\begin{figure}[tb]
	\centering
	\plotone{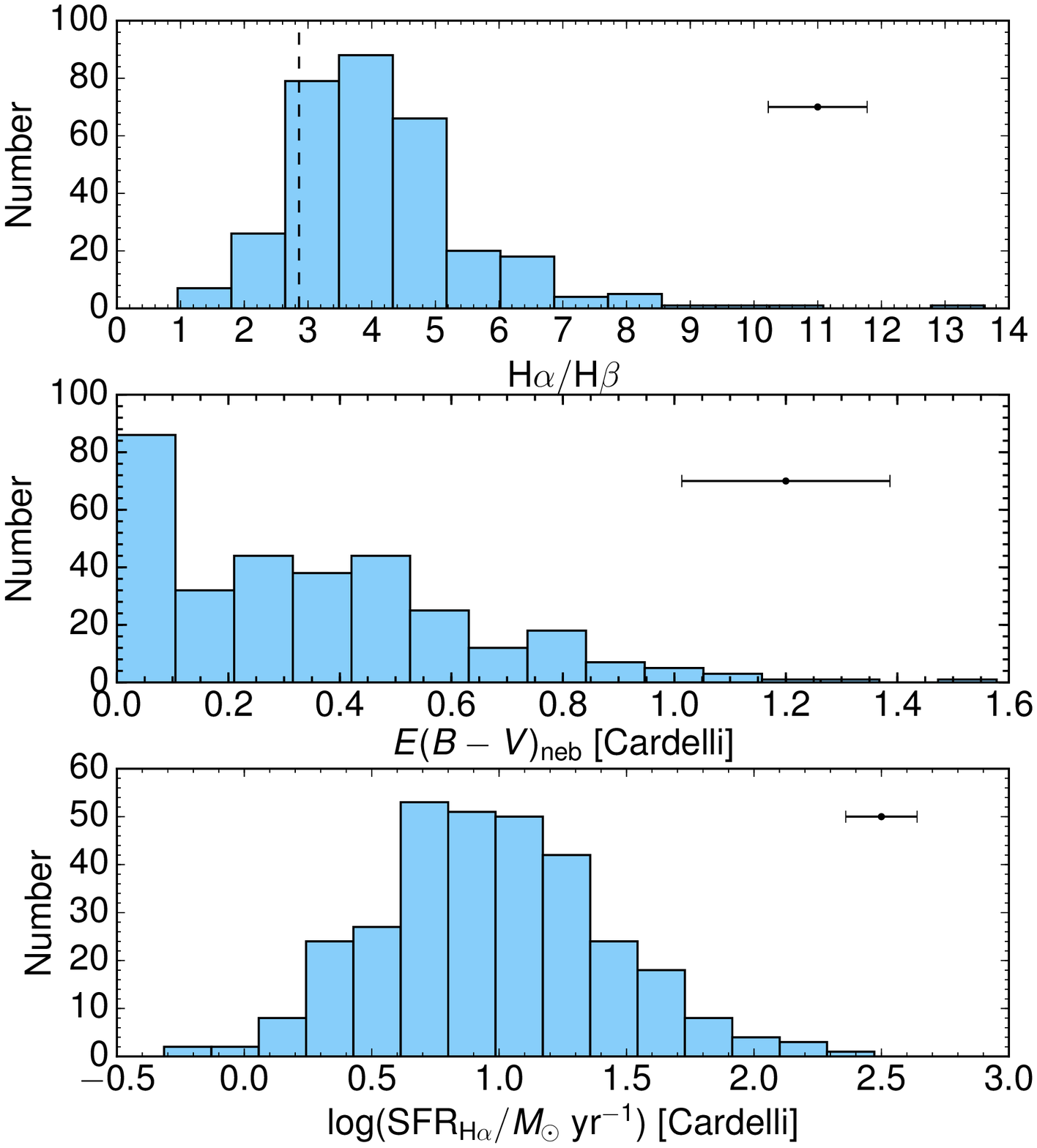}
	\caption{Histogram of \bd, \ebmvneb, and \sfrha\ for the 317 galaxies with $ > 3\sigma$ measurements of BD. The dashed line denotes the theoretical limit of BD in the case of no dust reddening, $\mathrm{BD} = 2.86$. Galaxies with $\mathrm{BD} \le 2.86$ (13\% of the sample) are assigned $E(B-V)_{\mathrm{neb}}=0$. Here \ebmvneb\ and \sfrha\ assume \citet{cardelli1989} extinction. The error bars represent the median uncertainty in each quantity. \label{fig:bdhist}}
	\end{figure}

Nebular reddening \ebmvneb\ was calculated from BD using the equation
\begin{equation}
	E(B-V)_{\mathrm{neb}} = \frac{2.5}{k(\mathHb) - k(\mathHa)}\log_{10}\left(\frac{\mathrm{BD}}{2.86}\right), \label{eqn:ebmv}
\end{equation}	
where $k(\mathHb)$ and $k(\mathHa)$ are the values of the attenuation curve evaluated at the wavelengths of \Hb\ and \Ha\ respectively; recall that $k(\lambda) = A_{\lambda}/E(B-V)$. 

The attenuation at \Ha\ is
\begin{equation}
\begin{aligned}
	A_{\mathHa} & = k(\mathHa)E(B-V)_{\mathrm{neb}} = 2.54E(B-V)_{\mathrm{neb}} \\
	& = 5.90\log_{10}\left(\frac{\mathrm{BD}}{2.86}\right)\label{eqn:aha_card}
\end{aligned}
\end{equation}
assuming \citet{cardelli1989} extinction and $R_{\mathrm{V}}=3.1$, or 
\begin{equation}
\begin{aligned}
	A_{\mathHa} & = k(\mathHa)E(B-V)_{\mathrm{neb}} = 2.17E(B-V)_{\mathrm{neb}} \\
	& = 4.95\log_{10}\left(\frac{\mathrm{BD}}{2.86}\right)\label{eqn:aha_smc}
\end{aligned}
\end{equation}
assuming SMC extinction and $R_{\mathrm{V}}=2.74$. The typical uncertainty on BD, accounting for the errors on the relative slit corrections in $H$ and $K$, is $14\%$. 
Of the 317 $z \sim 2.3$ galaxies studied in this paper, 51 (16\% of the sample) have BD $ < 2.86$, the nominal expectation for zero nebular reddening assuming Case B conditions and $T_{\mathrm{e}}=10,000$ K. All but 8 of these are consistent with BD $= 2.86$ within $2\sigma$. Galaxies with BD $\leq 2.86$ were assigned $E(B-V)_{\mathrm{neb}} = 0$.

SFRs (hereafter \sfrha) were estimated using the observed \Ha\ recombination line luminosities, corrected using Equation \ref{eqn:aha_card}, or Equation \ref{eqn:aha_smc} in cases where \sfrha\ was compared to a quantity where continuum attenuation assumes SMC. 

To determine the appropriate conversion between \Ha\ luminosity and SFR for each galaxy, we calculated the ionizing photon production rate corresponding to the best-fit SED by integrating the SED up to 912 \AA:
\begin{equation}
N({\mathrm{H}^0}) = \int_{0}^{912\text{\normalfont\AA}} \frac{\lambda f_{\lambda}}{hc} d\lambda.
\end{equation}
The BPASSv2.2 SEDs are in units of $f_{\lambda}$ per $M_{\odot}~\mathrm{yr}^{-1}$ of star formation, so the expected $L_{\mathrm{H}\alpha}$ per $M_{\odot}~\mathrm{yr}^{-1}$ of star formation can then be calculated using the equation from \citet{leitherer1995}:
\begin{equation}
L_{\mathrm{H}\alpha} [\mathrm{erg}~\mathrm{s}^{-1}] = \frac{1.36 N(\mathrm{H}^0)}{10^{12}} [\mathrm{s}^{-1}].
\end{equation}
For $t=10^8$ yr, assuming an IMF upper mass cutoff of $100M_{\odot}$:
\begin{equation}
\log(\mathrm{SFR}_{\mathrm{H}\alpha}/M_{\odot}~\mathrm{yr}^{-1}) = \log(L_{\mathrm{H}\alpha}/\mathrm{erg}~\mathrm{s}^{-1}) - 41.64\label{eqn:ha}
\end{equation}
For comparison, the constant value of 41.64 is 0.34 dex higher than the value of 41.30 from \citet{kennicutt2012}, once they have been converted to the same IMF. \Ha\ conversion factors for different model assumptions are given in Table \ref{tab:sfr}. The $L_{\mathrm{H}\alpha}/\mathrm{SFR}$ ratio depends on the spectral shape in the EUV, and when emission line ratios are used to identify models with plausible EUV spectral shapes \citep[as in][]{steidel2016} they are the models that produce significantly larger numbers of ionizing photons per $M_{\odot}$ of star formation. This issue is discussed in Section \ref{sec:sfr_disc}.

Also given in Table \ref{tab:sfr} are SFR calibrations evaluated at $\lambda = 1500$ \AA, as well as values of the ionizing photon production efficiency $\xi_{\mathrm{ion}}$. Although these numbers are not explicitly used in this paper, they are illustrative because \sfrsed\ is primarily sensitive to the non-ionizing UV luminosity, rather than the ionizing photon production rate, since it is derived entirely longward of 912 \AA. This can be seen in the fact that the 1500\AA\ SFR calibrations for the $300M_{\odot}$ and $100M_{\odot}$ BPASSv2.2 models are similar, since the non-ionizing UV luminosity per unit SF has very limited IMF dependence.

Figure \ref{fig:bdhist} shows the distribution of BD, \ebmvneb, and \sfrha, where \ebmvneb\ was calculated from BD assuming a \citet{cardelli1989} extinction curve. 

\subsection{Gas-phase oxygen abundances}
The values of gas-phase oxygen abundance quoted in this paper are inferred using the O3N2 index ($\mathrm{O3N2} \equiv \log[([\mathrm{O}~\textsc{iii}]\lambda5008/\mathrm{H}\beta)/([\mathrm{N}~\textsc{ii}]\lambda6585/\mathrm{H}\alpha)]$; \citealt{pp04}). \citet{steidel2014} showed that compared to the N2 index ($\mathrm{N2} \equiv \log([\mathrm{N}~\textsc{ii}]\lambda6585/\mathrm{H}\alpha)$; \citealt{pp04}), oxygen abundances measured from the O3N2 index were less affected by bias and scatter relative to direct $T_\mathrm{e}$-based measurements of individual KBSS-MOSFIRE galaxies. In this paper, we use the recalibration of O3N2 proposed by \citet{strom2017} based on the sample of extragalactic H~\textsc{ii} regions compiled by \citet{pilyugin2012}, with an 0.24 dex offset added to place the $T_e$ estimates (based on collisionally excited line ratios) onto the same scale as those measured from recombination lines  \citep{esteban2014,steidel2016}\footnote{This offset also makes $(\mathrm{N}/\mathrm{O}) = (\mathrm{N}/\mathrm{O})_{\odot}$ when $(\mathrm{O}/\mathrm{H}) = (\mathrm{O}/\mathrm{H})_{\odot}$ for the calibration sample.}:
\begin{equation}
 12 + \log(\mathrm{O}/\mathrm{H})_{\mathrm{O3N2}} = 8.80 - 0.20~\mathrm{O3N2}.\label{eqn:o3n2}
\end{equation}

Since $[\mathrm{O}~\textsc{iii}]\lambda5008$ and \Hb\ are close in wavelength and $[\mathrm{N}~\textsc{ii}]\lambda6585$ and \Ha\ are also close in wavelength, the ratio O3N2 and thus \Zneb\ is reddening-independent.
An important point made by several authors \citep[e.g.][]{steidel2014,sanders2016,strom2017} is that the ``strong-line'' metallicity calibrations established in the local universe may not be accurate at high redshift. The locus of the $z \sim 2.3$ KBSS-MOSFIRE sample is offset with respect to SDSS in the log([O~\textsc{iii}]$\lambda5008$/\Hb) versus log([N~\textsc{ii}]$\lambda6585$/\Ha) (N2-BPT) plane, likely driven by harder ionizing spectra at fixed nebular O/H \citep{steidel2014,steidel2016,strom2017}. Additionally, at high redshift, the strong-line ratios become less sensitive to the ionized gas-phase oxygen abundance and more sensitive to the overall spectral shape of the ionizing radiation field produced by massive stars. In practice, this means that the inference of gas-phase metallicity from the O3N2 index is uncertain at high redshift, since O3N2 also tracks excitation. We continue to use O3N2-based inferences of 12 + log(O/H) in this paper, with the caveat that trends of increasing \Zneb\ may also be interpreted as trends of decreasing excitation.
    
\begin{figure*}[tbh]
\centering
\epsscale{0.75}
\plotone{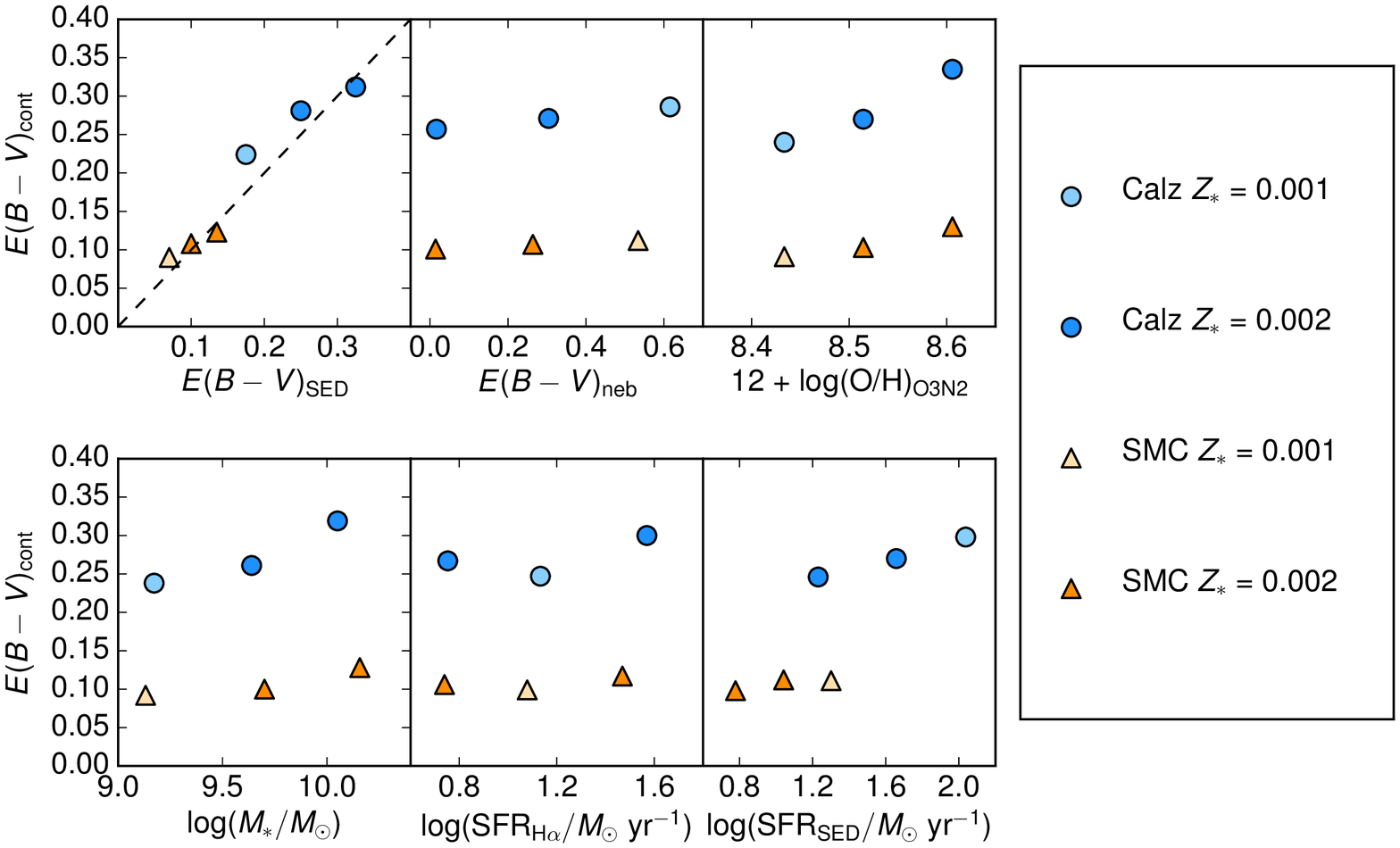}
\caption{Trends between \ebmvcont\ as measured from the stacked LRIS spectra shown in Figure \ref{fig:uvfits} and the six binned quantities: \ebmvsed, BD (here converted to \ebmvneb), \Zneb, $M_*$, \sfrha, and \sfrsed. The blue circles correspond to model fits assuming the \citet{calzetti2000} attenuation curve and the orange triangles assume SMC. In both cases, the light-colored points correspond to the spectra best-fit by a $Z_* = 0.001$ model, and the dark-colored points correspond to those best-fit by a $Z_* = 0.002$ model. While \ebmvcont\ shows increasing trends with every quantity, the stellar metallicity of the best-fit model is consistently low and is not correlated with any of the binned quantities. \label{fig:trends}}
\end{figure*}
	
\section{Inferences from composite rest-UV spectra}\label{sec:rest-uv}

Composite rest-UV spectra were generated for the subset of 270 galaxies out of 317 with complementary LRIS and MOSFIRE spectra described in Table \ref{tab:sample} (hereafter the LRIS+MOSFIRE sample). To fit model spectra to the observed data, we used the far-UV spectra generated by the BPASSv2.2 model suite \citep{stanway2018} assuming a constant star formation history with an age of $10^8$ yr, an upper mass IMF cutoff of $100M_{\odot}$, and binary evolution included. The stellar metallicity $Z_*$ and reddening \ebmvcont\ were free parameters in the fitting procedure, and the best-fit combination of $Z_*$ and \ebmvcont\ was determined by a $\chi^2$ minimization. Details of the fitting procedure are described in Appendix \ref{sec:fitting}.

Figure \ref{fig:trends} shows the best-fit values of \ebmvcont\ as a function of each of the six binned quantities: \ebmvsed, \bd\ (here converted to \ebmvneb), \Zneb, \mstar, \sfrha, and \sfrsed. Results are shown for both \citet{calzetti2000} and SMC attenuation.

The bins in \ebmvsed\ served as a test case, to ensure that photometric and spectroscopic fitting results were internally consistent. We expect \ebmvsed\ and \ebmvcont\ to be strongly correlated, although the correlation might not be 1:1 due to the fact that the stellar metallicity corresponding to \ebmvcont\ is allowed to vary freely, whereas the SED fitting fixes the stellar metallicity at $Z_* = 0.002$\footnote{We expect $Z_* \sim 0.001-0.002$; see below.}. The best-fit values of \ebmvcont\ increase monotonically with increasing \ebmvsed\ as expected and stay very close to the median values of \ebmvsed\ in the bins. 

We find that all composites are best fit by either a $Z_* = 0.001$ model ($Z_*/Z_{\odot} = 0.07$) or a $Z_* = 0.002$ model ($Z_*/Z_{\odot} = 0.14$). This is as we might expect based on the analysis from \citet{steidel2016} which used a similar $\chi^2$ minimization analysis for the ``LM1'' stack (which contains some of the same galaxy spectra as this sample) and found that models with $Z_* \geq 0.003$ were strongly dis-favored compared to those with $Z_* \leq 0.002$. Similarly, \citet{strom2017,strom2018} showed using photoionization modeling that only binary models with low $Z_*$ were able to reproduce the nebular line ratios observed in most individual $z \sim 2.3$ galaxies.

Since the ages we infer for the galaxies in our sample are $\sim50-300$ Myr, and the delay timescale for Fe enrichment of the ISM by Type Ia supernovae is $\sim 300-400$ Myr, the ISM metallicities in these galaxies (and thus the abundances of the massive stars) are likely dominated by core-collape supernovae. Rising star formation histories, which are expected for galaxies during the period of most rapid growth \citep{reddy2012a}, can extend the time period over which ISM enrichment remains dominated by core-collapse SNe. \citet{steidel2016} argued that under these conditons, galaxies can reach near-solar gas-phase O/H while Fe abundance remains low. Our result is consistent with this general picture, where stellar metallicity is uniformly low at all values of nebular reddening (Figure \ref{fig:uvfits}b) and \Zneb\ (Figure \ref{fig:uvfits}c).

\begin{figure*}[!htb]
	\centering
    \plottwo{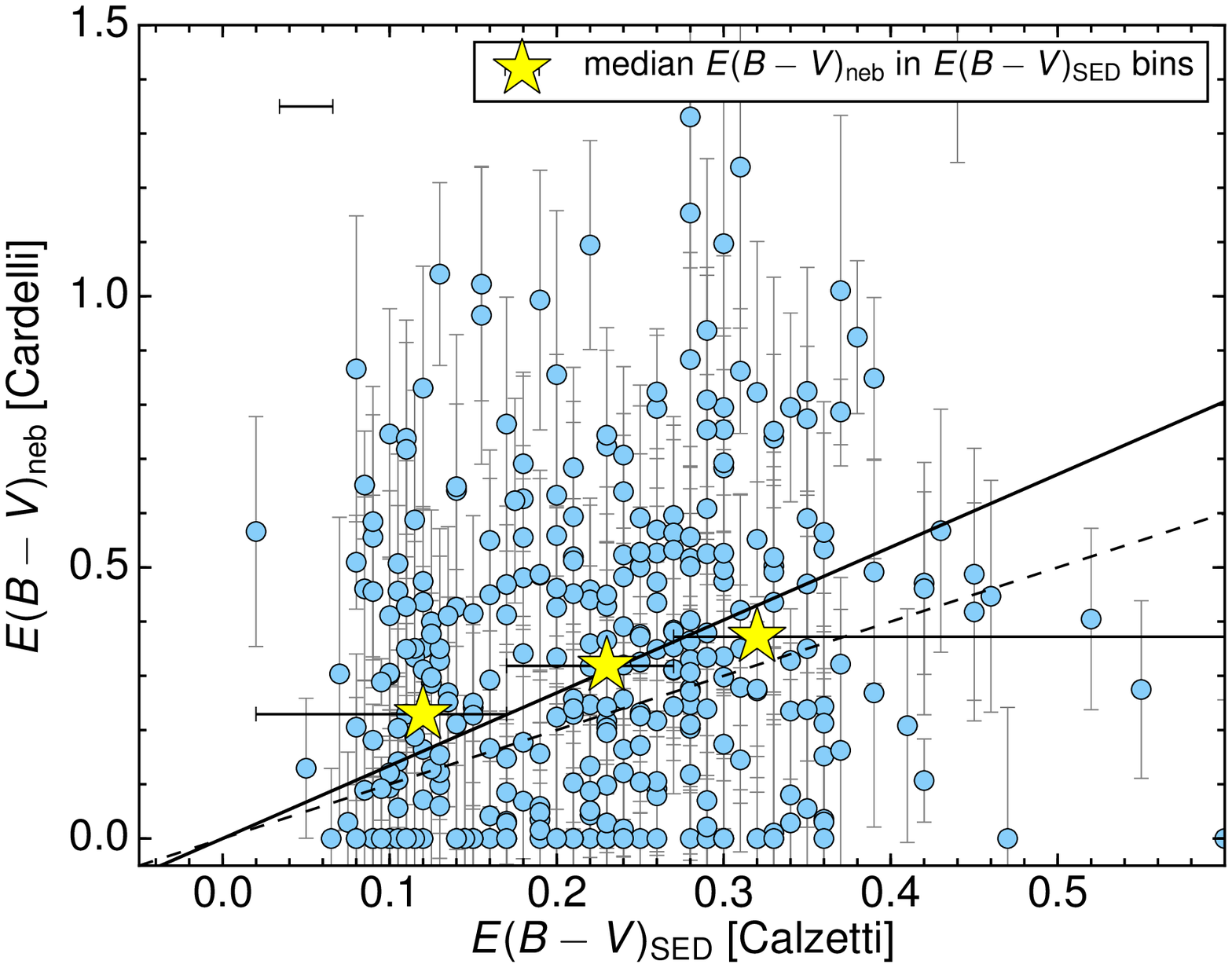}{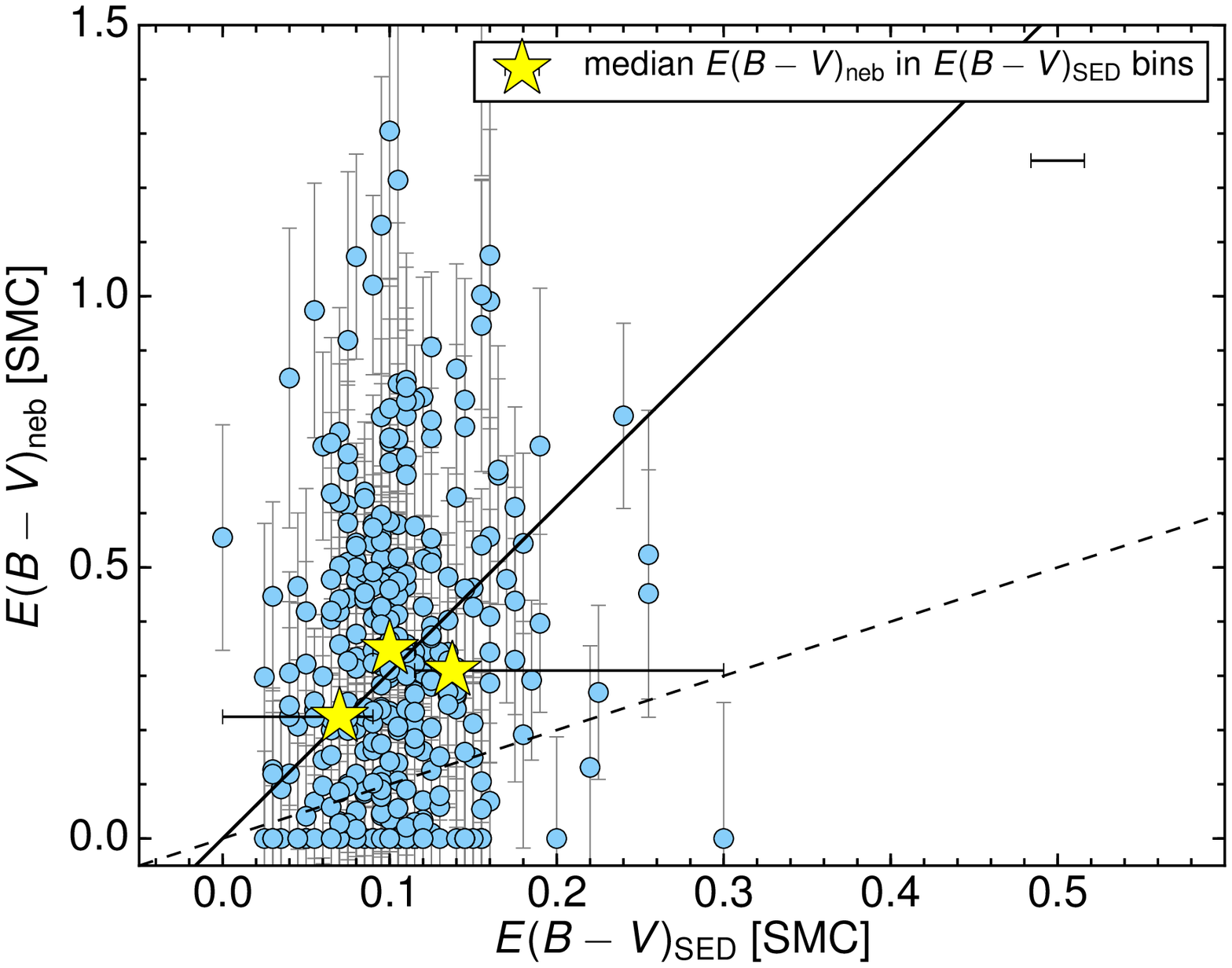}
	\caption{\ebmvneb\ versus \ebmvsed. The solid lines represent orthogonal distance regressions fit to the data (Equation \ref{eqn:ebmvcomp}), similar to \citet{calzetti2000}, and the dashed lines represent equal \ebmvneb\ and \ebmvsed. Yellow stars are median values of \ebmvneb\ in equal-number bins of \ebmvsed, and the bin limits are represented as horizontal error bars. The error bars on \ebmvneb\ correspond to a $1\sigma$ error in \bd, and the horizontal error bar is the representative error on \ebmvsed\ discussed in Appendix \ref{sec:sed}. \emph{Left}: The SED fitting assumes a \citet{calzetti2000} attenuation curve and \ebmvneb\ assumes a \citet{cardelli1989} line-of-sight extinction curve. \emph{Right}: Both attenuation estimates assume SMC extinction. For both assumed curves, \ebmvneb\ is generally greater than \ebmvsed, with large scatter; however, the apparent discrepancy between the two reddening estimates is greater when the SMC curve is assumed. \label{fig:ebmv}}
	\end{figure*}

While \ebmvcont\ is positively correlated with all six binned quantities, the strength of the correlation varies---the weakest trend is with \ebmvneb. In the following section, we explore these relationships for \emph{individual galaxies} using \ebmvsed, which we have argued is an equally good measure of continuum reddening---we show that the same general trends still hold.

\section{Relationship between nebular and continuum attenuation}\label{sec:ebmv_comp}

Figure \ref{fig:ebmv} compares \ebmvsed\ to \ebmvneb for two combinations of continuum attenuation curve and nebular extinction curve. The solid lines in Figure \ref{fig:ebmv} represent orthogonal distance regressions fit to the data, which can be used to derive a translation between \ebmvsed\ and \ebmvneb, similar to that proposed by \citet{calzetti2000}. With this sample we find:
\begin{align}\label{eqn:ebmvcomp}
E(B-V)_{\mathrm{neb, MW}} &= (1.34 \pm 0.05)E(B-V)_{\mathrm{SED, Calz}} \\
E(B-V)_{\mathrm{neb, SMC}} &= (3.06 \pm 0.15)E(B-V)_{\mathrm{SED, SMC}}.
\end{align}
The error bars on the coefficients are the standard deviation of the orthogonal distance regression, and the median scatter about each the best-fit lines (in the \ebmvneb\ direction) are 0.20 (Cardelli/Calzetti) and 0.19 (SMC). Spearman rank-order correlation coefficients for both combinations of curves are given in Table \ref{tab:corr}.

\begin{figure*}[!htb]
	\centering
	\plottwo{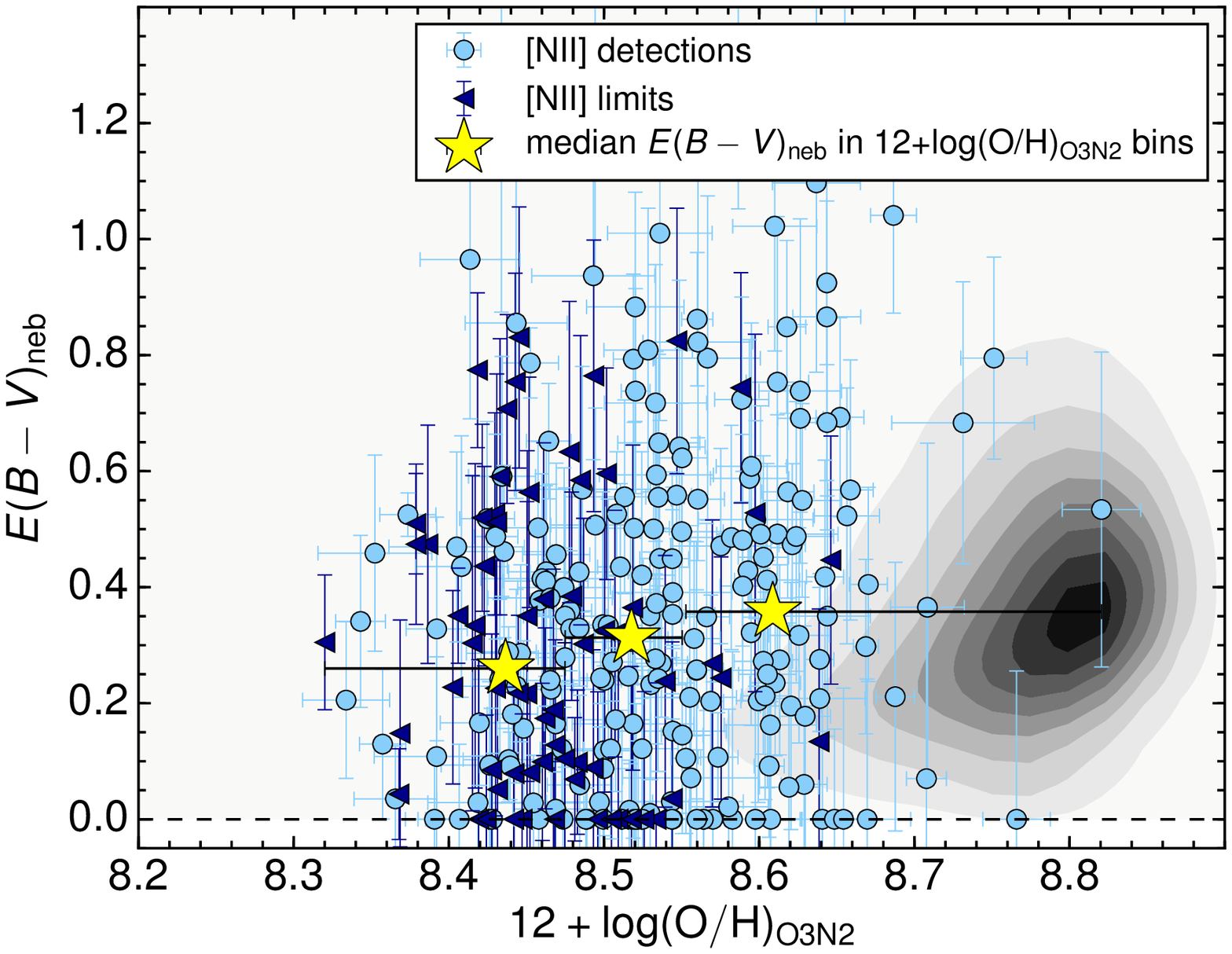}{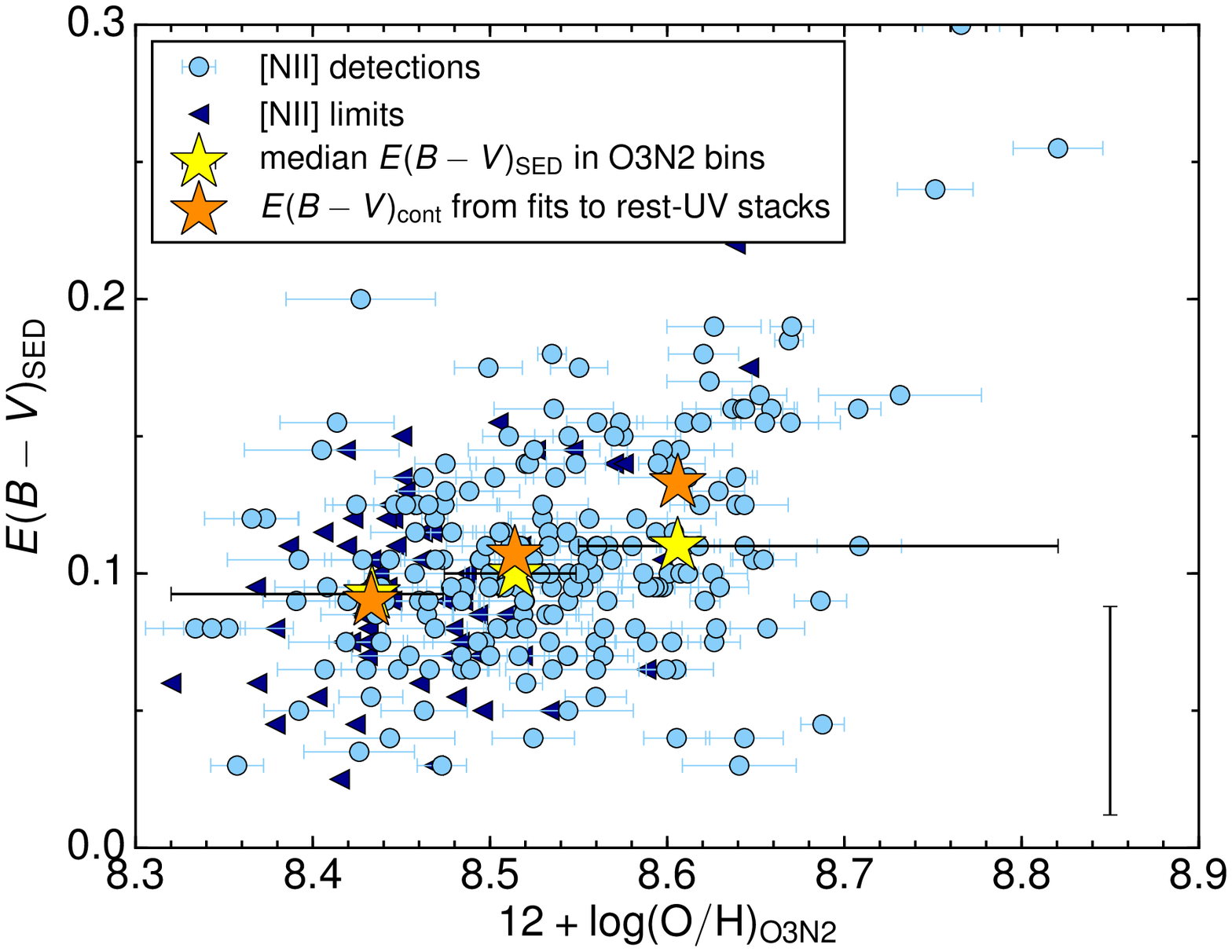}
	\caption{\emph{Left:} \ebmvneb\ versus \Zneb. Dark blue triangles represent $2\sigma$ upper limits in [N~\textsc{ii}].  Yellow stars represent the median measurements of the individual spectra in four equal-number bins of \Zneb. The grayscale contours represent the locus of similarly selected SDSS galaxies \citep{strom2017}. \emph{Right:} \ebmvsed\ versus \Zneb, assuming the SMC attenuation curve. Orange stars are measurements of \ebmvcont\ for stacks of LRIS spectra in bins of \Zneb, and yellow stars are median values of \ebmvsed in the same bins. Bin limits are shown as black error bars. While we find no significant correlation between \Zneb\ and \ebmvneb, \Zneb\ and \ebmvsed\ are correlated at $3.0\sigma$ significance. Additionally, KBSS galaxies have a lower \Zneb\ than SDSS galaxies by 0.21 dex on average. \label{fig:ebmv_o3n2}}
\end{figure*}

\begin{deluxetable*}{lDD}
\centering
\tablecaption{Spearman Rank-Order Correlations\label{tab:corr}}
\tablehead{\colhead{Correlated Quantities\tablenotemark{a}} & \multicolumn2c{Spearman $\rho$\tablenotemark{b}} & \multicolumn2c{Spearman $\sigma$\tablenotemark{c}}} 
\decimals
\startdata
\ebmvneb\ [Cardelli] vs. \ebmvsed\ [Calzetti] & 0.15 & 2.6 \\
\ebmvneb\ [SMC] vs. \ebmvsed\ [SMC]           & 0.13 & 2.3 \\
\Zneb\ vs. \ebmvneb\ [Cardelli]               & 0.06 & 1.1 \\
\Zneb\ vs. \ebmvsed\ [Calzetti]               & 0.17 & 3.0 \\
\Zneb\ vs. \ebmvsed\ [SMC]                    & 0.17 & 3.0 \\
\ebmvneb\ [Cardelli] vs. $M_*$ [Calzetti]     & 0.23 & 4.2 \\
\ebmvneb\ [SMC] vs. $M_*$ [SMC]               & 0.21 & 3.7 \\
\sfrha\ [Cardelli] vs. \sfrsed\ [Calzetti]    & 0.24 & 4.3 \\
\sfrha\ [SMC] vs. \sfrsed\ [SMC]              & 0.33 & 6.0 \\
\enddata
\tablenotetext{a}{The assumed attenuation curves are indicated in brackets.} 
\tablenotetext{b}{Spearman rank correlation coefficient.} \tablenotetext{c}{Standard deviations from null hypothesis.}
\end{deluxetable*}

\subsection{Relationship between reddening and gas-phase metallicity}\label{sec:o3n2_bd}

One might expect dust reddening and gas-phase enrichment to be strongly correlated, as metals in the ISM will be depleted onto the same dust grains that attenuate the nebular emission lines---the higher the metal content of the ISM, the more dust grains will be formed \citep{reddy2010}. Motivated by this picture, the lefthand panel of Figure \ref{fig:ebmv_o3n2} compares measurements of \ebmvneb\ to \Zneb. The lefthand panel of Figure \ref{fig:ebmv_o3n2} shows, for comparison, low-redshift galaxies from SDSS-DR7 \citep[][grayscale contours]{abazajian2009}, selected to have similar detection properties as KBSS-MOSFIRE \citep[see][]{strom2017}. Figure \ref{fig:ebmv_o3n2} shows that compared to SDSS, the KBSS sample has systematically lower \Zneb\ by 0.21 dex on average at a given value of \ebmvneb. The righthand panel of Figure \ref{fig:ebmv_o3n2} compares \Zneb\ with \ebmvsed, where the orange stars are measurements of \ebmvcont\ for the stacked rest-UV spectra in bins of \Zneb (see Section \ref{sec:rest-uv}). The measurements of \ebmvcont\ for the stacks show good agreement with the median \ebmvsed\ in the bins (yellow stars), again demonstrating that these quantities on average provide consistent estimates of the stellar continuum reddening.

We observe no statistically significant correlation between \Zneb\ and \ebmvneb\ ($1.1\sigma$; Table \ref{tab:corr}). However, we do observe a strong correlation between \Zneb\ and \ebmvsed\ ($3.0\sigma$). 

\begin{figure*}[!htb]
\centering
\plottwo{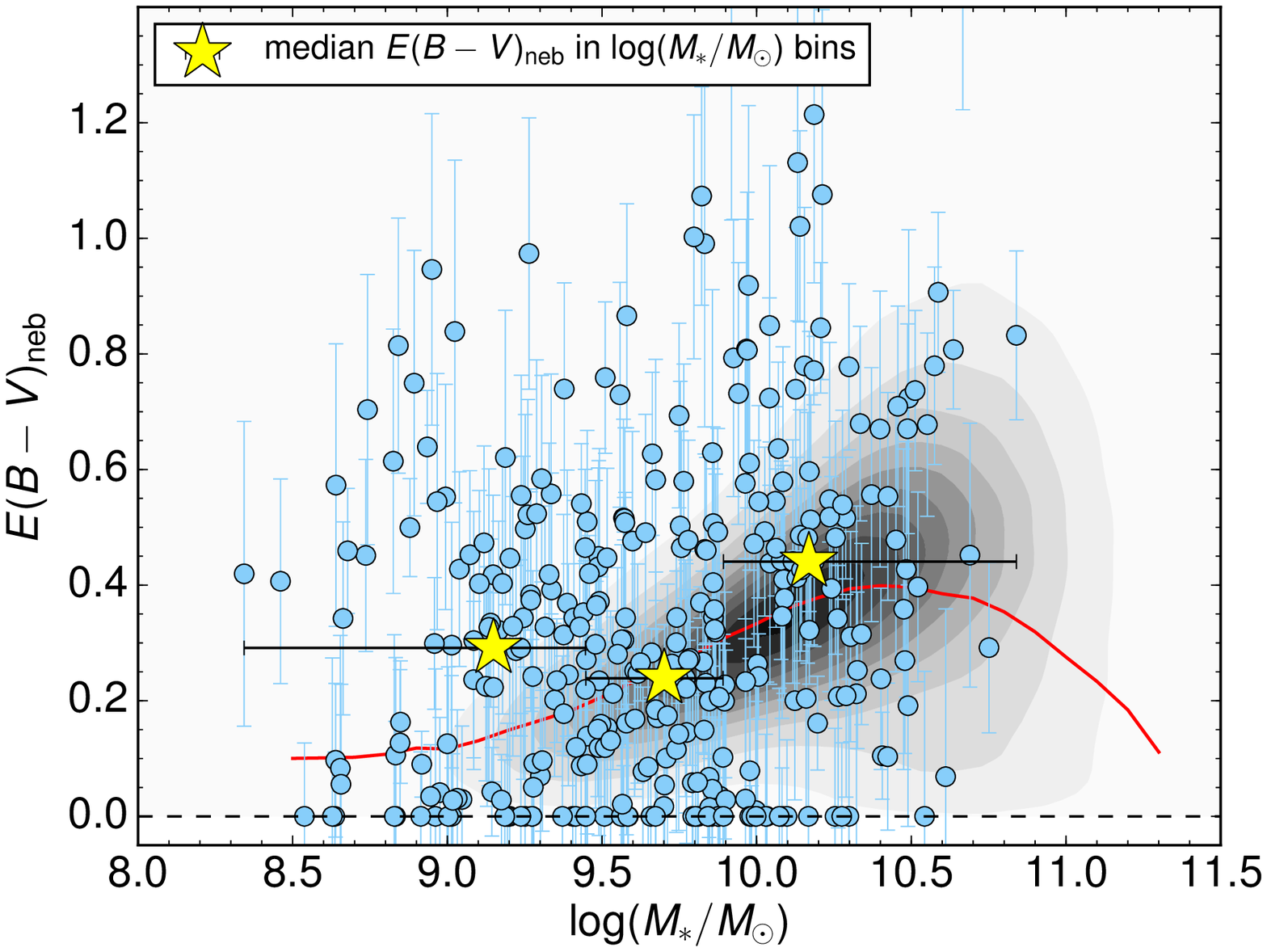}{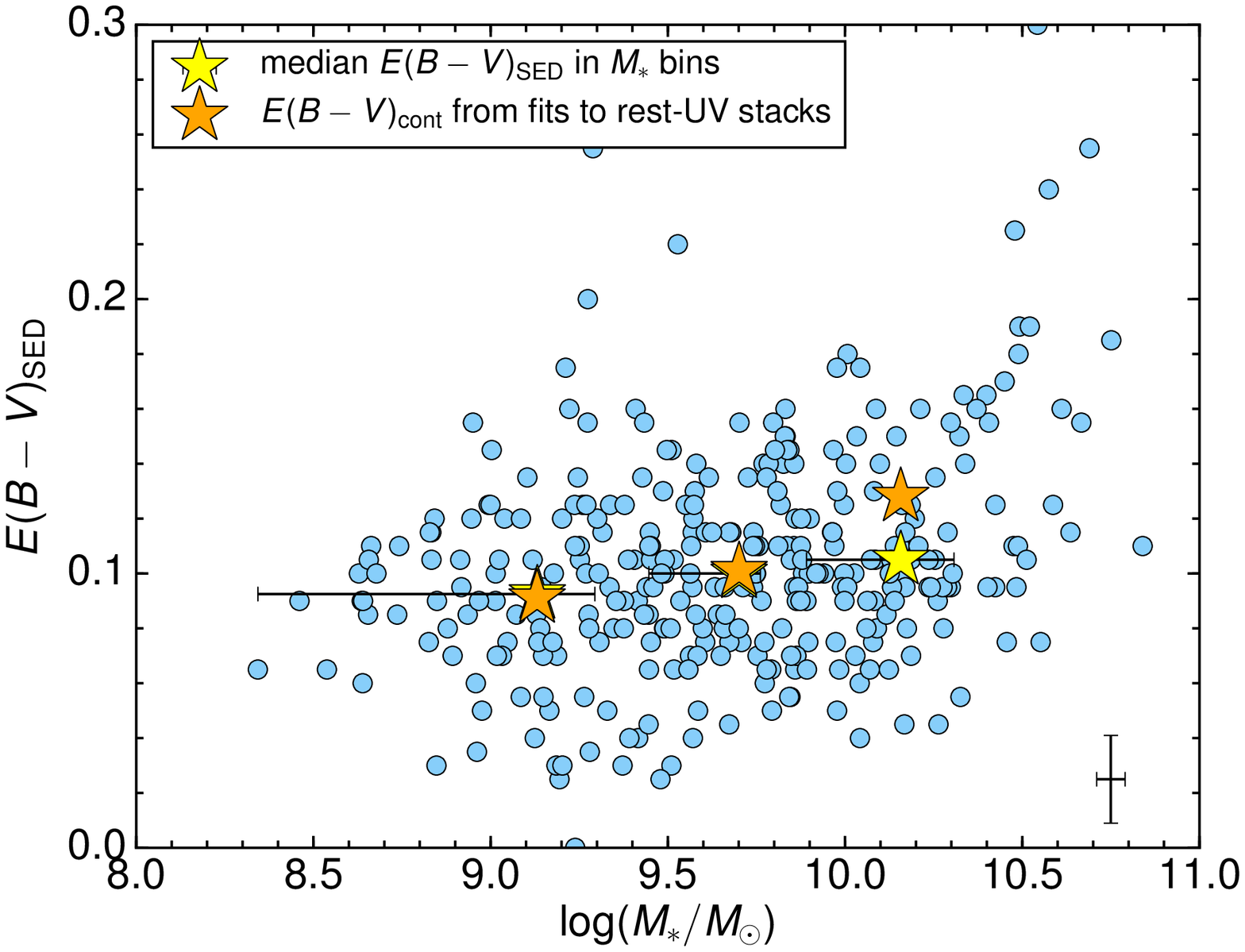}
\caption{\emph{Left:} \ebmvneb\ versus $M_*$ inferred from SED fits assuming a BPASSv2.2 $Z_* = 0.002$ continuous star formation model with SMC attenuation. The solid red line represents the locus of SDSS galaxies, determined by the median \ebmvneb\ in bins of \mstar. The KBSS and SDSS galaxies are consistent with being drawn from the same parent distribution in \ebmvneb, but KBSS galaxies are offset to lower stellar masses; thus, the degree of reddening at fixed stellar mass is higher on average in KBSS galaxies. \emph{Right:} \ebmvsed\ versus $M_*$, where both quantities have been inferred from SED fits assuming a BPASSv2.2 $Z_* = 0.002$ continuous star formation model with SMC attenuation. The yellow stars represent the median values of \ebmvsed\ in equal-number bins of $M_*$, and the orange stars are measurements of \ebmvcont\ from stacks of LRIS spectra in the same bins. Bin limits are shown as black error bars. As in Figure \ref{fig:ebmv_o3n2}, the trend with continuum reddening are stronger than the trend with nebular reddening \label{fig:mstar_ebmv}}
\end{figure*}

\subsection{Relationship between $E(B-V)$, \mstar, and SFR}

The lefthand panel of Figure \ref{fig:mstar_ebmv} compares measurements of \ebmvneb\ to \mstar, where local galaxies from SDSS-DR7 are shown for comparison. The median stellar mass of the KBSS sample is 0.72 dex lower than that of the SDSS sample, and a Kolmogorov-Smirnov (K-S) test revealed that the two samples are drawn from different parent distributions in \mstar. However, they are consistent with being drawn from the same parent distribution in \ebmvneb. Both samples show a positive correlation between \ebmvneb\ and \mstar. Thus, at fixed \mstar, KBSS galaxies have a greater degree of nebular reddening on average.

The righthand panel of Figure \ref{fig:mstar_ebmv} compares \ebmvsed\ to \mstar; while these quantities may be degenerate due to the fact that they are both derived from the SED fits, the presence of a correlation is supported by the correlation between \ebmvcont\ and \mstar\ (orange stars), which are measured independently. This may be explained by the following argument. \mstar\ is related to the normalization of the model SED relative to the data. \ebmvsed\ essentially parametrizes the ``shape'' of the SED. While the whole spectrum is used for normalization, the matching of the overall shape of the SED is likely to be driven primarily by the shape in the UV, which is also sensitive to parameters such as age and star formation history. Since higher Fe/H at fixed O/H will make UV spectra intrinsically redder for older galaxies, dust reddening and maturity of star formation could be degenerate.

\begin{figure*}[!htb]
	\centering
    \plottwo{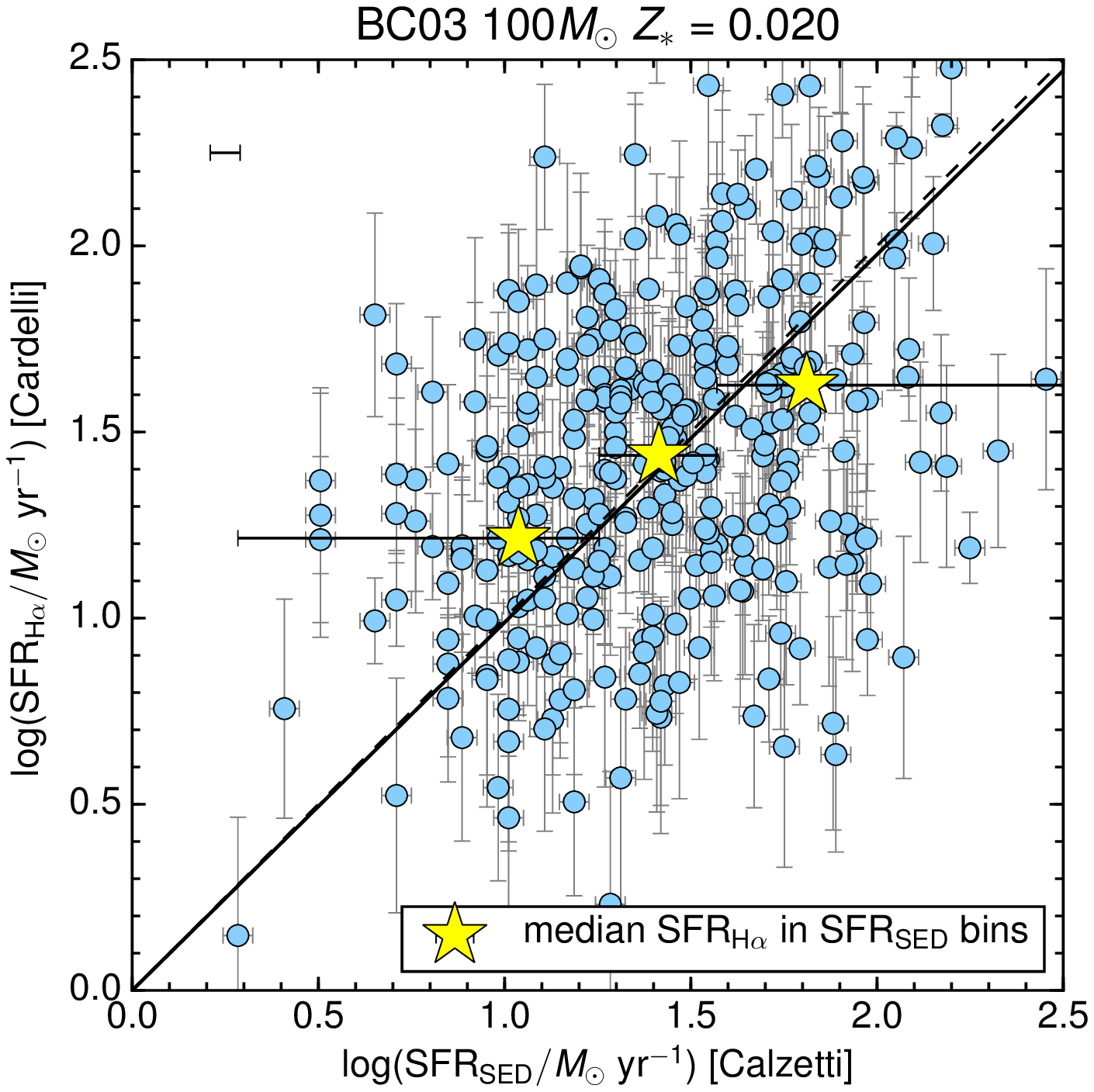}{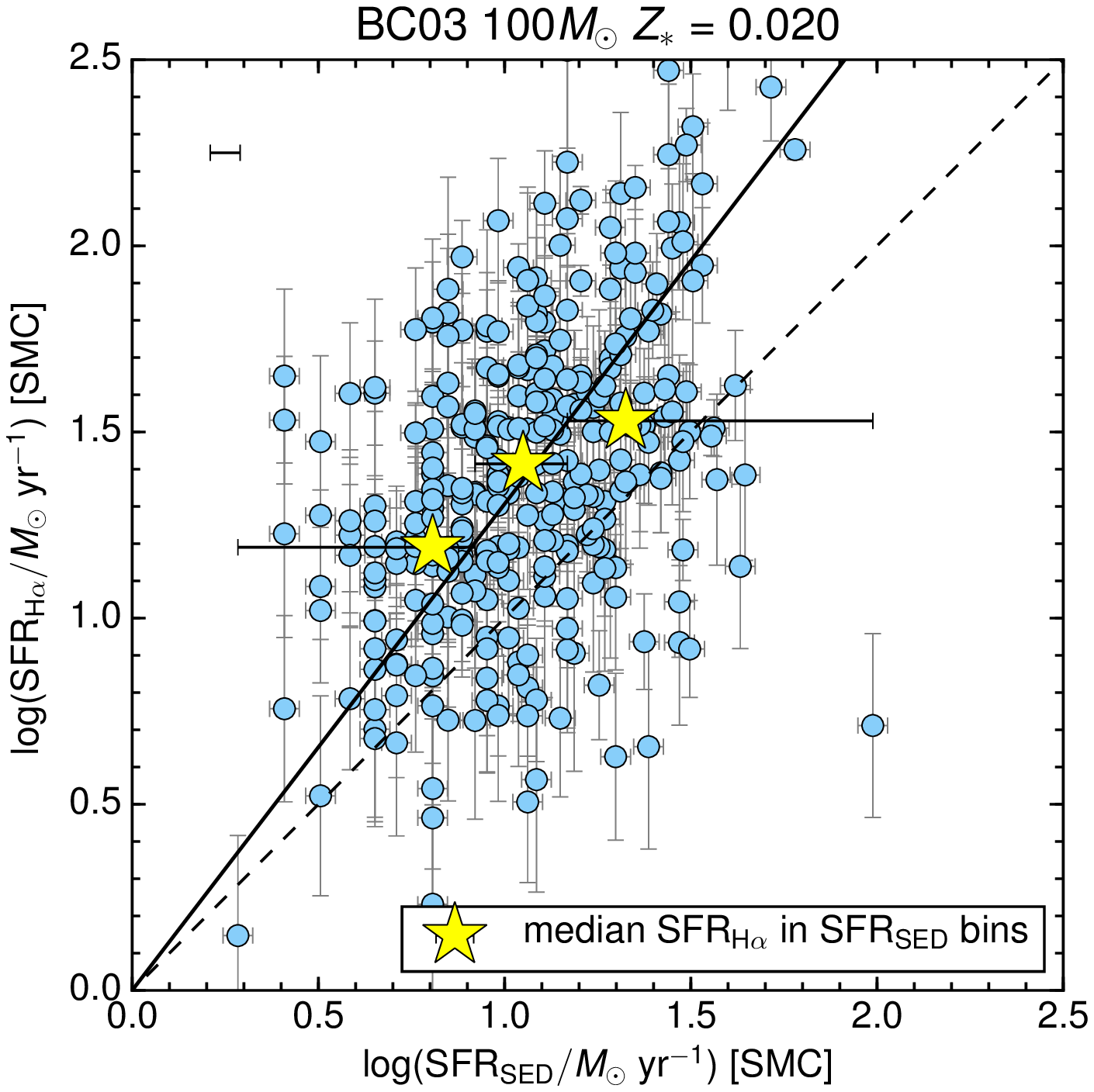}
	\caption{Comparison of star formation rates estimated from SED fitting (\sfrsed) with those based on the \Ha\ luminosity (\sfrha), for the \citet{bruzual2003} model with $Z_*=0.020$}. The solid lines are orthogonal distance regression lines, and the dashed lines represent equal SFRs. Yellow stars are median vales of \sfrha\ in equal-number bins of \sfrsed, and bin limits are represented as horizontal error bars. \emph{Left}: The SED fitting assumes a \citet{calzetti2000} attenuation curve and the \Ha\ measurements assume a \citet{cardelli1989} line-of-sight extinction curve. \emph{Right}: Both the SED fitting and the \Ha\ measurements assume SMC extinction. Using the SMC extinction curve results in values of $\log(\mathrm{SFR}_{\mathrm{SED}}$ that are offset from $\log(\mathrm{SFR}_{\mathrm{H}\alpha}$ by a factor of 1.3, whereas using \citet{cardelli1989} extinction and \citet{calzetti2000} attenuation results in consistent SFRs on average. \label{fig:sfr_comp_bc03}
	\end{figure*}

\begin{figure*}[!htb]
	\centering
    \plottwo{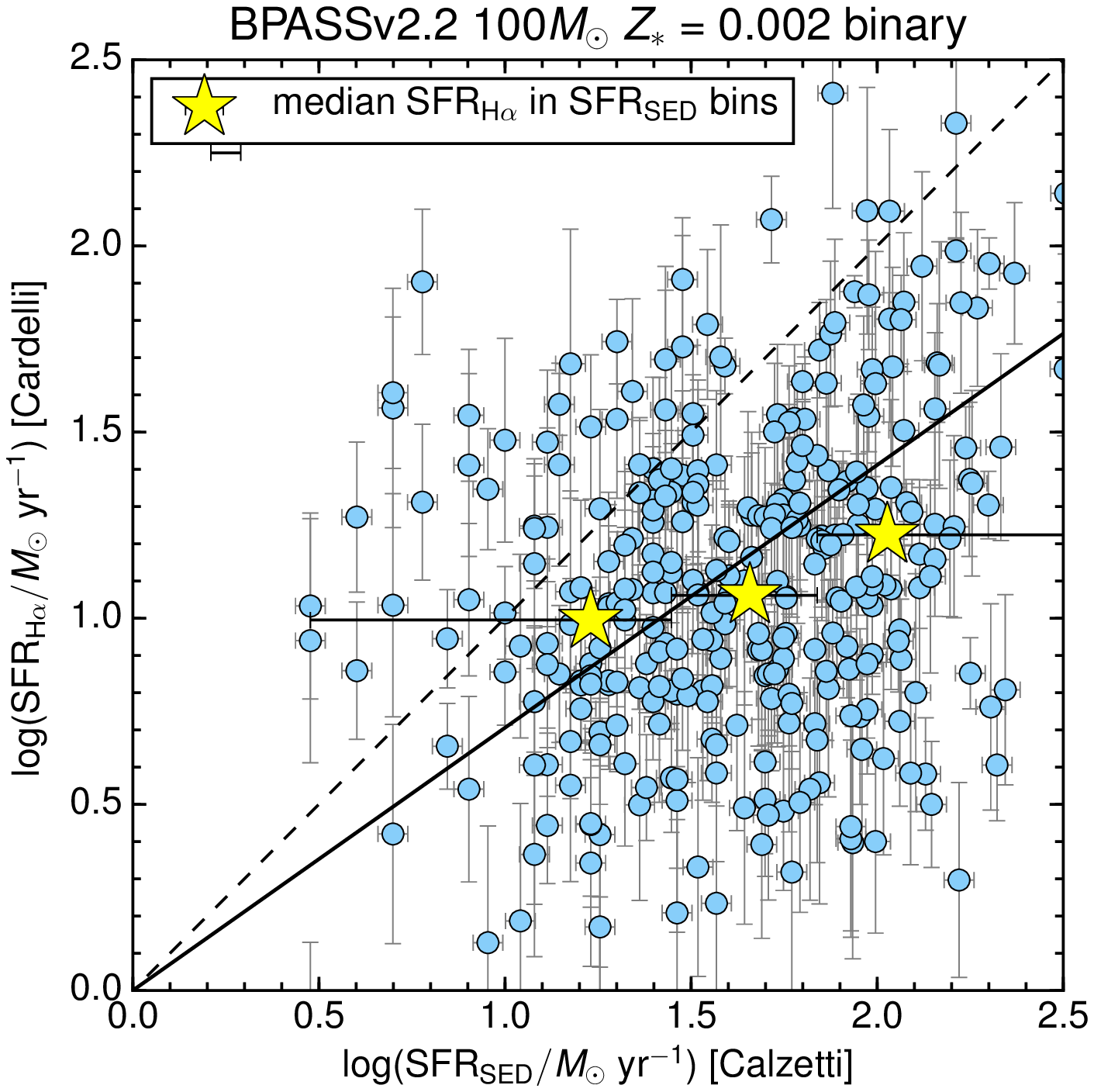}{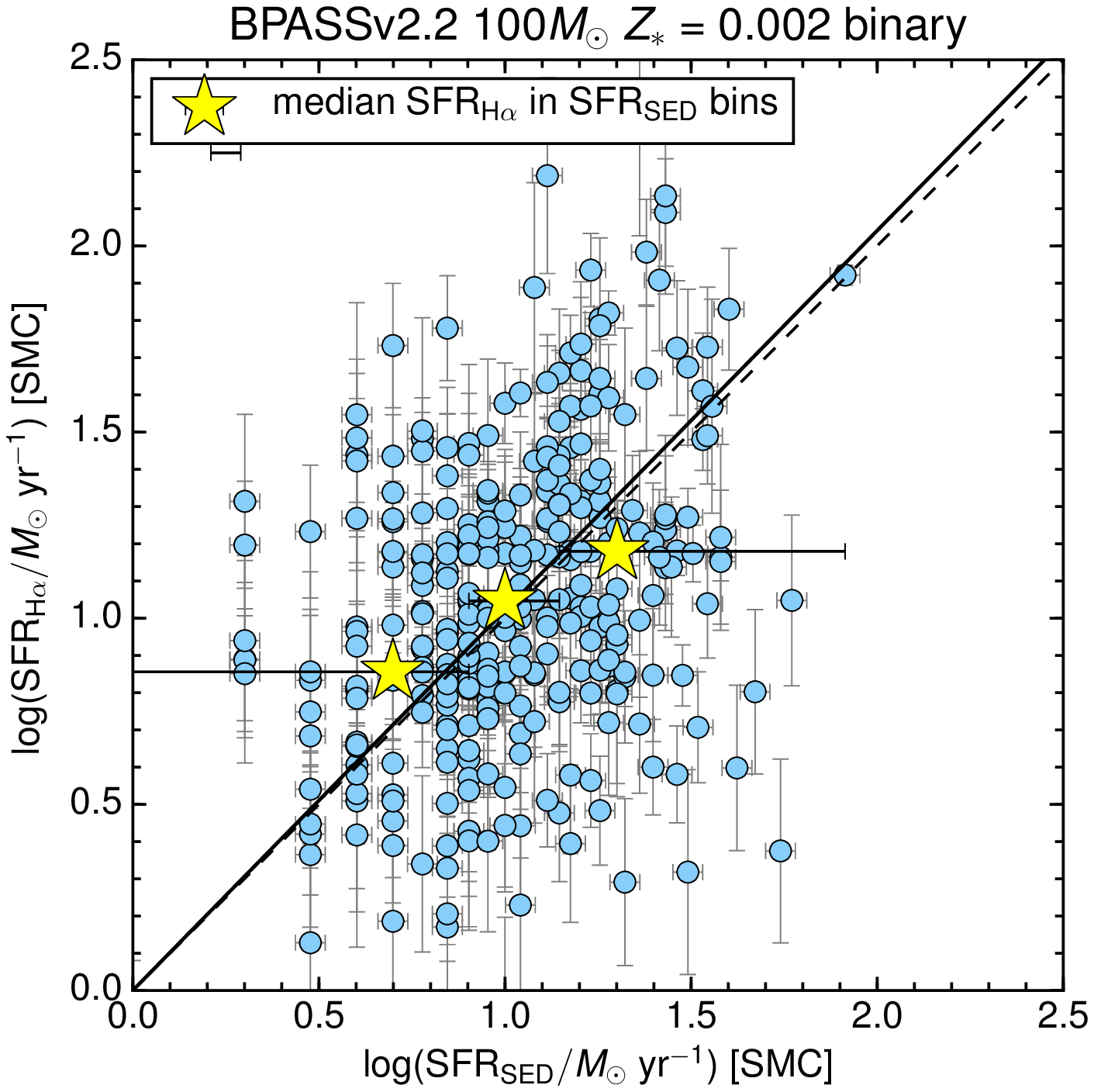}
	\caption{Comparison of star formation rates estimated from SED fitting (\sfrsed) with those based on the \Ha\ luminosity (\sfrha). The solid lines are orthogonal distance regression lines, and the dashed lines represent equal SFRs. Yellow stars are median vales of \sfrha\ in equal-number bins of \sfrsed, and bin limits are represented as horizontal error bars. \emph{Left}: The SED fitting assumes a \citet{calzetti2000} attenuation curve and the \Ha\ measurements assume a \citet{cardelli1989} line-of-sight extinction curve. \emph{Right}: Both the SED fitting and the \Ha\ measurements assume SMC extinction. Using the \citet{calzetti2000} curve results in values of $\log(\mathrm{SFR}_{\mathrm{SED}}$ that are offset from $\log(\mathrm{SFR}_{\mathrm{H}\alpha}$ by a factor of 1.3, whereas using the SMC curve results in consistent SFRs on average. \label{fig:sfr_comp}}
	\end{figure*}
	
\begin{figure}[!htb]
\centering
\plotone{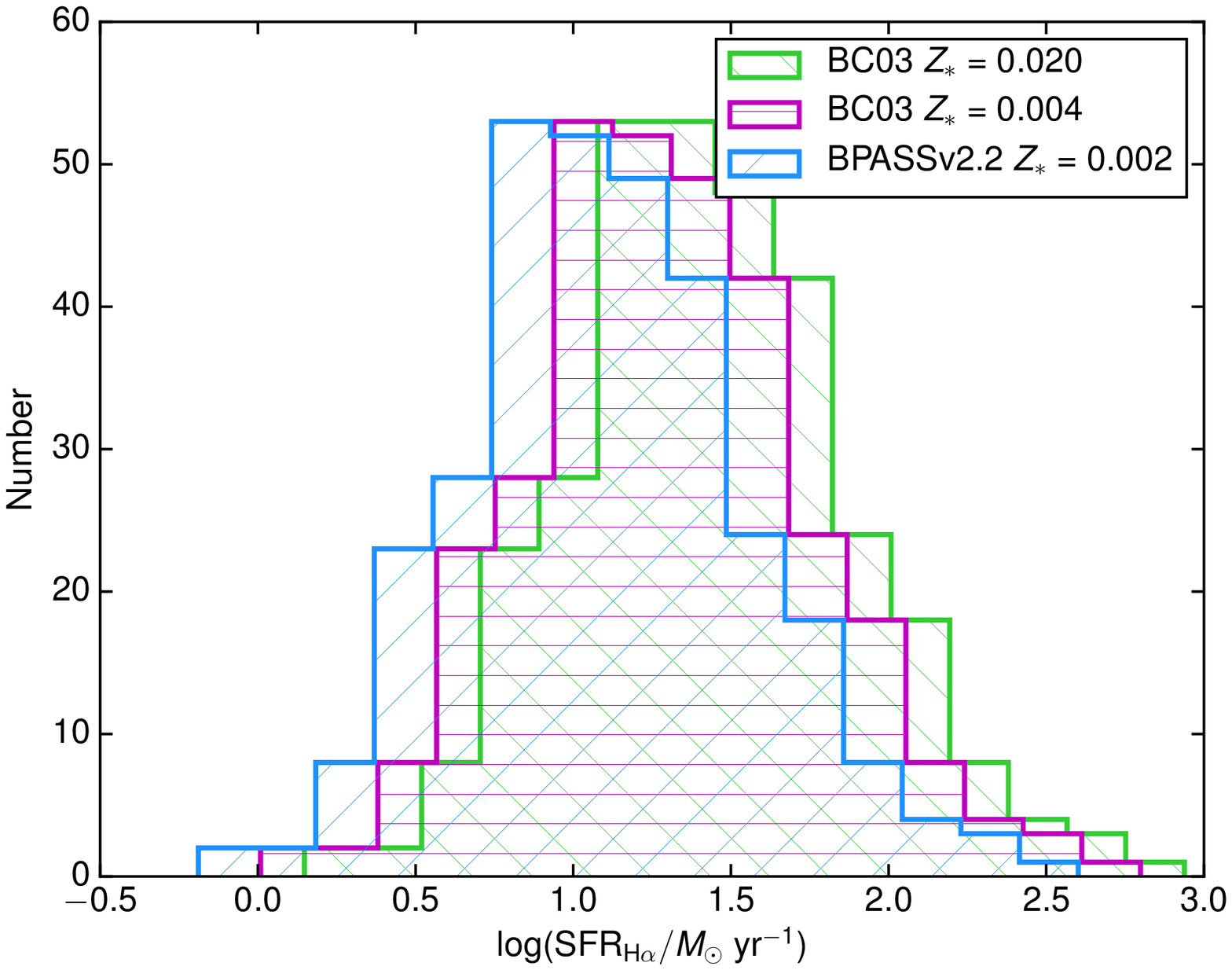}
\caption{Histograms of \sfrha\ for BPASSv2.2 $100 M_{\odot}$ $Z_*=0.002$ binary models (blue), \citetalias{bruzual2003} $Z_*=0.004$ (magenta), and \citetalias{bruzual2003} $Z_*=0.020$ (green). In all cases, the default BPASS IMF is assumed, and \Ha\ luminosities have been converted to SFRs by integrating the ionizing portion of the best-fit SED, as described in Section \ref{sec:neb}. Compared to BPASSv2.2, the \citetalias{bruzual2003} $Z_*=0.004$ models predict values of \sfrha\ that are 0.27 dex higher on average, and the \citetalias{bruzual2003} $Z_*=0.020$ models predict values of \sfrha\ that are 0.5 dex higher. \label{fig:sfr_hist}}
\end{figure}

\section{SFR comparisons}\label{sec:sfr_comp}

In this paper, SFRs are estimated from both SED fitting (which is primarily sensitive to the UV continuum) and \Ha\ luminosities. If the correct curve is assumed in both the UV and optical regime and all systematic uncertainties have been accounted for, then these two measures of SFR should agree on average in the case of continuous star formation.

Figure \ref{fig:sfr_comp_bc03} compares \sfrha\ to \sfrsed\ for the \citet{bruzual2003} $Z_*=0.020$ models, assuming \citet{calzetti2000} attenuation and \citet{cardelli1989} extinction (left), and SMC for both lines and continuum (right). Also shown in Figure \ref{fig:sfr_comp} are orthogonal distance regression lines fit to the data. Using the SMC extinction curve results in values of $\log(\mathrm{SFR}_{\mathrm{SED}}$ that are offset from $\log(\mathrm{SFR}_{\mathrm{H}\alpha}$ by a factor of 1.3, whereas using \citet{cardelli1989} extinction and \citet{calzetti2000} attenuation results in consistent SFRs on average.

Conversely, Figure \ref{fig:sfr_comp} compares the same quantities, but instead for the BPASSv2.2 $Z=0.002$ binary models. In this case, using the \citet{calzetti2000} curve results in values of $\log(\mathrm{SFR}_{\mathrm{SED}}$ that are offset from $\log(\mathrm{SFR}_{\mathrm{H}\alpha}$ by a factor of 1.3, whereas using the SMC curve results in consistent SFRs on average.

Figure \ref{fig:sfr_hist} compares the distributions of \sfrha\ and \sfrsed\ for BPASSv2.2 $100 M_{\odot}$ $Z_*=0.002$ binary models, \citetalias{bruzual2003} $Z_*=0.004$, and \citetalias{bruzual2003} $Z_*=0.020$. In all cases, the default BPASS IMF has been assumed, and \Ha\ luminosities have been converted to SFRs by integrating the ionizing portion of the best-fit SED, as described in Section \ref{sec:neb}. Compared to BPASSv2.2, the \citetalias{bruzual2003} $Z_*=0.004$ models predict values of \sfrha\ that are 0.20 dex higher on average, and the \citetalias{bruzual2003} $Z_*=0.020$ models predict values of \sfrha\ that are 0.5 dex higher on average, once the differences in IMF have been accounted for. \sfrsed\ assuming the BPASS models is higher than that predicted by \citetalias{bruzual2003} $Z=0.020$ by 0.12 dex, also accounting for differences in IMF.

On the other hand, when the same SED model is assumed, changing the continuum attenuation curve from \citet{calzetti2000} to SMC lowers \sfrsed\ by 0.69 dex on average, and changing the nebular extinction curve from \citet{cardelli1989} to SMC lowers \sfrha\ by 0.06 dex on average. Thus, changing the combination of assumptions from \citetalias{bruzual2003} $Z_*=0.020$ models with \citet{calzetti2000} attenuation and \citet{cardelli1989} extinction to BPASSv2.2 $Z_*=0.002$ models with SMC applied to both lines and continuum lowers \sfrsed\ by 0.57 dex and \sfrha\ by 0.56 dex. This can be seen clearly in Figures \ref{fig:sfr_comp_bc03} and \ref{fig:sfr_comp}; differences in attenuation curve cancel out differences in SED model. In order to break this degeneracy and determine the most applicable attenuation relation for our sample, independent constraints on the best-fit SED model are needed.

\section{Discussion}\label{sec:disc}

\subsection{Comparison between nebular and continuum reddening}\label{sec:ebmv_comp_disc}

Figure \ref{fig:ebmv} shows that although a correlation exists between the reddening toward the ionized nebulae and the stellar continuum, there is large scatter. In particular, galaxies with $E(B-V)_{\mathrm{neb}} = 0$ (which is defined here as $\mathrm{BD} \leq 2.86$) span the full range of \ebmvsed. Similarly, the measurements of \ebmvcont\ for stacked spectra in bins of BD show only a very weak trend, suggesting that nebular reddening is nearly independent of UV continuum reddening. 

This relationship between \ebmvneb\ and \ebmvcont\ is intriguing because previous studies have reached different conclusions. For example, \citet{steidel2016} found that the best-fit \ebmvcont\ for the LM1 stack was consistent with the median \ebmvneb\ of the individual galaxies contributing to the stack. They suggested that the similarity between these two quantities implies that the nebular emission lines are powered by the same massive stars that are responsible for the far-UV continuum, and thus there is not likely to be a significant population of dust-obscured massive stars contributing to the nebular emission line luminosities but not the far-UV continuum. We find that with the \citet{calzetti2000} attenuation curve and \citet{cardelli1989} extinction curve, $E(B-V)_{\mathrm{neb}} = 1.34 E(B-V)_{\mathrm{SED}}$ on average, in contrast to the factor of 2.27 between \ebmvneb\ and \ebmvcont\ proposed by \citet{calzetti2000} for local starburst galaxies. When the SMC curve is applied to both lines and continuum, we find that $E(B-V)_{\mathrm{neb}} = 3.06 E(B-V)_{\mathrm{SED}}$, which is merely due to the steeper slope of the SMC curve in the UV relative to the \citet{calzetti2000} curve.

At low redshift, the discrepancy between nebular and continuum color excess is usually attributed to increased dust covering fractions surrounding H~\textsc{ii} regions; i.e., the youngest stars remain in undissipated parent birth clouds \citep{calzetti1994,charlot2000}. However, this picture does not explain the significant fraction of galaxies in the KBSS sample for which $E(B-V)_{\mathrm{SED}} > E(B-V)_{\mathrm{neb}}$. 

Additionally, scatter in the relationship between \ebmvneb\ and \ebmvcont\ may be due to variations in the dust attenuation curve from galaxy to galaxy. It is unlikely that the same attenuation curve applies to every galaxy in our sample, or even every galaxy in a bin of some observed quantity. Theoretical work has suggested that observed variations in the IRX-$\beta$ relation may be attributed to variations in the dust attenuation curve due to differences in grain composition, as well as stellar population age, dust temperature, and geometry of the dust distribution \citep[e.g.][]{granato2000,popping2017,narayanan2017}. Indeed, we find that the same curve cannot be applied to every galaxy in our sample. While 38\% of galaxies in our sample have SEDs that are best fit by SMC, 16\% are best fit by \citet{calzetti2000}, and the remainder do not favor either curve within the uncertainty (although the majority of these are nominally better fit by SMC).

Regardless of which curve is assumed, however, the scatter in \ebmvneb\ relative to \ebmvsed\ is large. Similarly, Figures \ref{fig:ebmv_o3n2} and \ref{fig:mstar_ebmv} show that while \ebmvsed\ is strongly correlated with both \Zneb\ and \mstar, neither quantity is strongly correlated with \ebmvneb.

\subsection{Correlations with gas-phase metallicity and stellar mass}\label{sec:Zneb_disc}

Figure \ref{fig:ebmv_o3n2} shows that local galaxies have a higher gas-phase metallicity (and/or lower excitation) at the same \ebmvneb\ than high-redshift galaxies. This is consistent with the well-studied offset in the mass-metallicity relation \citep[MZR;][]{tremonti2004,erb2006,mannucci2010,zahid2011,steidel2014} at high redshift, such that high-redshift galaxies have lower metallicity at fixed \mstar.

A number of studies \citep[e.g.][]{reddy2010,heinis2014,oteo2014,alvarez-marquez2016,bouwens2016} have noted correlations between continuum dust obscuration and \mstar; we find such a correlation for the KBSS sample as well (Figure \ref{fig:mstar_ebmv}). Other studies have found a redshift-independent relationship between \emph{nebular} attenuation and stellar mass; \citet{dominguez2013} found using stacked WFC3 grism spectra of galaxies at intermediate redshift ($0.75 \leq z \leq 1.5$) that these galaxies have a similar BD at fixed stellar mass as SDSS galaxies (within the errors). However, we find that our sample of $z\sim2.3$ galaxies has a distribution of BD that is similar to that of SDSS galaxies (albeit with large scatter), but a median \mstar\ that is 0.6 dex lower than that of $z\sim0$ SDSS galaxies. In both samples there is a positive correlation between \ebmvneb\ and \mstar. Thus, \ebmvneb\ is higher in KBSS galaxies than in SDSS galaxies at a fixed \mstar. The presence of a decrement in the MZR but not in the dust obscuration-stellar mass relation may be explained by the expectation that if dust and metals are produced in the same way across redshift, high-redshift galaxies should have lower dust-to-gas ratios than low-redshift galaxies at the same stellar mass; however, their larger gas fractions \citep[e.g.][]{daddi2008,tacconi2010} overcompensate for this to produce a larger column of dust at fixed stellar mass. It may also be the case that dust grain composition changes systematically with redshift.

Interestingly, we found (Section \ref{sec:o3n2_bd}) that the correlation between stellar continuum reddening \ebmvcont\ and (reddening-independent) \Zneb\ is much stronger than that between nebular reddening \ebmvneb\ and \Zneb\ (Figure \ref{fig:ebmv_o3n2}; Table \ref{tab:corr}). This may be explained if the intrinsic scatter is much larger for \ebmvneb, a conclusion supported by the weak correlation between \ebmvneb\ and \ebmvsed\ discussed in the previous section. Alternately, this trend could be explained if \ebmvsed\ is a better tracer of gas-phase abundances and/or excitation. Several authors \citep[e.g.][]{steidel2014,sanders2016,strom2017} have noted that at high redshift, line ratios such as O3N2 become less sensitive to gas-phase oxygen abundance and more sensitive to the overall spectral shape of the ionizing radiation field produced by massive stars. Thus, the stronger correlation between \Zneb\ and \ebmvsed\ may be induced because both quantities are closely tied to the spectral shape of the ionizing radiation field produced by the massive stars, which depends on the stellar abundance of Fe.

\subsection{SFR comparisons}\label{sec:sfr_disc}

Figures \ref{fig:sfr_comp} and \ref{fig:sfr_comp_bc03} show that the combination of \citet{calzetti2000} attenuation for the stellar continuum and \citet{cardelli1989} Galactic extinction for the nebular emission lines produces consistent \Ha-based and SED-based SFRs when the \citetalias{bruzual2003} $Z=0.020$ single-star models are used, as in \citet{steidel2014}. When the BPASSv2.2 $Z=0.002$ binary models (and appropriate \Ha\ conversions) are used, SMC produces consistent SFRs.

As mentioned in Section \ref{sec:curves}, the SEDs of 46\% of the sample do not significantly favor either \citet{calzetti2000} or SMC within the uncertainty. Similarly, while the BPASS models provide a better fit than \citetalias{bruzual2003} for the majority of the sample, 33\% do not significantly favor either model within the uncertainty. Thus, the SED fitting alone is not able to constrain either the most appropriate dust attenuation curve or the most appropriate SPS model. Indeed, Figure \ref{fig:sfr_comp_bc03} appears to indicate that the combination of \citet{calzetti2000} attenuation, \citet{cardelli1989} extinction, and the single-star, solar metallicity \citetalias{bruzual2003} models is a reasonable combination of assumptions, and this is the combination of assumptions used by nearly all previous studies.

However, the BPASSv2.2 $Z_*=0.002$ models are more consistent with observations than the $Z_*=0.020$ \citetalias{bruzual2003} models in that they are better able to simultaneously match the rest-UV continuum, stellar and nebular lines, and rest-optical nebular emission lines. Thus, we argue that the agreement between SFRs when the single-star models are used is likely to be coincidental rather than an indication that these models are an accurate represention of the conditions at high redshift. When the BPASSv2.2 $Z_*=0.002$ model and corresponding \Ha-to-SFR conversion is used, SMC produces consistent SFRs on average, and so we argue that it is the most appropriate attenuation curve for the majority of KBSS galaxies, in agreement with what has been found in several recent studies \citep[e.g.,][]{capak2015,bouwens2016,koprowski2016,reddy2017}.

The large scatter in \sfrha\ with respect to \sfrsed\ could be explained by several factors. First, it is likely that the same \Ha/SFR conversion factor does not apply to every galaxy in our sample. The $L_{\mathrm{H}\alpha}/L_{\mathrm{UV}}$ ratio is directly proportional to the ionizing photon production efficiency $\xi_{\mathrm{ion}}$. \citet{shivaei2017} found an intrinsic scatter of 0.28 dex in the distribution of $\xi_{\mathrm{ion}}$ for galaxies in the MOSDEF survey. They conclude that variations in $\xi_{\mathrm{ion}}$ cannot be solely explained by object-to-object variations in the dust attenuation curve; rather, the scatter is affected by stellar population properties such as variations in IMF and stellar metallicity. The conversion from \Ha\ luminosity to SFR predicted by the BPASSv2.2 models is directly calculated from the model predictions for $\xi_{\mathrm{ion}}$ for a given model template, but it is likely that the intrinsic $\xi_{\mathrm{ion}}$ of a given galaxy varies from the model predictions. Thus, variations in $\xi_{\mathrm{ion}}$ could explain the scatter in the \sfrsed/\sfrha\ ratio from galaxy to galaxy. 

One potential caveat is that variations between \Ha\ and UV-based SFRs are frequently attributed to  ``bursty'' star formation \citep[e.g.][]{weisz2012,kauffmann2014,smit2016,shivaei2016}. It is often asserted that the \Ha\ flux traces SFR over $\approx 10$ Myr timescales and the UV continuum flux traces SFR over $\approx 200$ Myr timescales \citep{kennicutt2012}, in which case a galaxy with a heightened \Ha/UV flux ratio may have had a short burst of star formation within the last 10 Myr \citep{sparre2017}. This effect could explain some of the scatter in Figure \ref{fig:sfr_comp}, given that \sfrsed\ is predominantly determined by the UV spectrum and its slope. However, binary stars can significantly extend the timescales over which ionizing photons are produced \citep{eldridge2017}, and for the galaxies in our sample, we do not expect significant fluctuations in SFR on timescales shorter than $\sim 30$ Myr, the typical central dynamical timescale; note also that the SEDs and FUV spectra are well-fit by models with constant star formation histories with a typical timescale of $10^8$ yr. Even in a scenario where bursts were important, we might expect to see a \sfrha/\sfrsed\ ratio that was systematically elevated, which is not the case, at least for the fiducial BPASS model employed in this paper.

\begin{figure}
\centering
\plotone{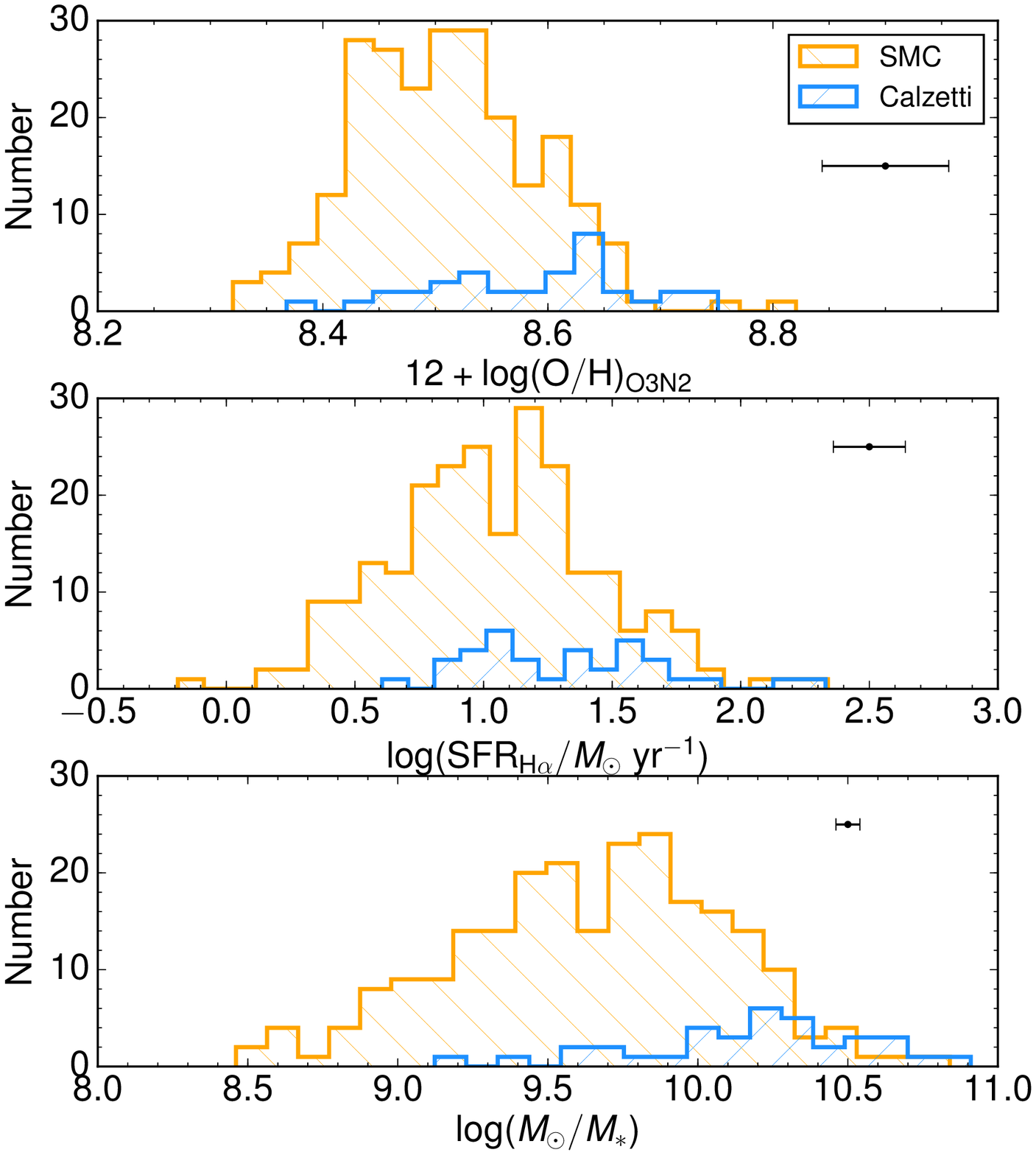}
\caption{Histograms of \Zneb, \sfrha, and \mstar\ for the 16\% of galaxies with SEDs best fit by the \citet{calzetti2000} curve (blue) and the remainder of the sample (orange). Note that the former two quantities are independent of the assumed continuum attenuation curve, and we have shown that while the latter quantity requires an assumption of attenuation curve, it is nearly independent of this choice (see Figure \ref{fig:mstar}).\label{fig:calz_hist}}
\end{figure}

Additionally, while the SMC curve comes the closest to producing consistent SFRs, and it provides the best fit to the SEDs of many of the galaxies in the sample, it may not be the most appropriate attenuation curve for every galaxy. As discussed in Section \ref{sec:curves}, 16\% of galaxies in the sample are better fit by \citet{calzetti2000} within the uncertainties.

Interestingly, this subset of galaxies better fit by the \citet{calzetti2000} curve displays unique properties. A Kolmogorov-Smirnov (KS) test indicates that these galaxies and the remainder of the sample are drawn from different parent distributions in \Zneb\ ($3.2\sigma$), \sfrha\ ($2.8\sigma$), and \mstar\ ($7.0\sigma$). Histograms of these three properties for the subsample best fit by \citet{calzetti2000} and the remainder of the sample are shown in Figure \ref{fig:calz_hist}. Thus, it may be the case that this subsample is a population of galaxies distinct from the majority.

\section{Summary and Conclusions}\label{sec:summary}

We have presented an analysis of a sample of high-quality near-IR spectra of 317 galaxies at $2.0 \leq z \leq 2.7$ obtained as a part of KBSS-MOSFIRE \citep{steidel2014,strom2017} in combination with complementary deep optical spectra obtained with Keck/LRIS for 270 of these galaxies. We have estimated \ebmvneb\ from the Balmer decrement and continuum attenuation from SED fits to broadband photometry [\ebmvsed] as well as from stellar spectral synthesis model fits to composite rest-UV spectra [\ebmvcont] in bins of \ebmvneb, \Zneb, $M_*$, \sfrha, and \sfrsed. We compared nebular and continuum estimates of dust reddening to each other and to gas-phase metallicity, $M_*$, and SFR. Finally, we compared \Ha\ and SED-based estimates of SFR. Our conclusions are as follows:
\begin{itemize}

\item \ebmvneb\ is correlated with \ebmvsed\ for individual galaxies, albeit with large scatter, and there is generally larger reddening towards line-emitting regions. Fits of BPASSv2.2 stellar spectral synthesis models to composite rest-UV spectra in bins of \ebmvneb\ confirm that \ebmvcont\ is also correlated with \ebmvneb. We find that when the SMC curve is applied to both lines and continuum, the discrepancy between \ebmvneb\ and \ebmvsed\ is larger on average than when the \citet{calzetti2000} curve is applied to the continuum and the \citet{cardelli1989} curve is applied to the nebular lines, due to the fact that the SMC curve is much steeper than the \citet{calzetti2000} curve in the UV but is similar to the \citet{cardelli1989} Milky Way extinction curve in the optical. We argue that translations between nebular and continuum reddening such as that proposed by \citet{calzetti2000} should be used with caution given the large scatter.

\item  \ebmvneb\ is correlated with gas-phase O/H measured from the O3N2 index, which is also sensitive to the massive stars ionizing the nebulae \citep[e.g.][]{steidel2014,steidel2016,strom2017}. The sense of the observed correlation is that the nebular reddening increases with increasing gas-phase O/H and/or decreasing excitation in the H~\textsc{ii} regions; the trend is similar to that observed at low redshift, but is offset by $\Delta {\rm log(O/H)} = 0.21$ dex toward lower inferred gas-phase oxygen abundance. Similarly, the relationship between \ebmvneb\ and $M_*$ is offset such that $z\sim2.3$ galaxies have greater nebular reddening at a fixed stellar mass than $z \sim 0$ galaxies. We interpret the behavior as a natural consequence of lower dust/gas ratios but much higher gas fractions at high redshift.

\item We find that the {\it continuum} reddening measured from the far-UV spectra is more strongly correlated with the inferred ionized gas-phase O/H (measured using the dust-insensitive O3N2 index) than the nebular reddening measured from the Balmer decrement toward the same \ion{H}{2} regions. The strength of this correlation may be due to the fact that both the O3N2 index, used to measure \Zneb, and the UV spectral slope, used to measure \ebmvcont, are sensitive to the shape of the ionizing radiation field produced by the massive stars.

\item The use of the \citet{calzetti2000} attenuation curve in SED fitting in combination with the BPASSv2.2 $Z_* = 0.002$ SED models and corresponding conversion from \Ha\ luminosity to SFR produces inconsistent values of \sfrsed\ and \sfrha\ for all but the highest-\sfrha\ galaxies when the BPASSv2.2 $Z=0.002$ binary star models are assumed, which several authors have found is best able to reproduce the observed constraints at high redshift. In contrast, the SMC curve produces consistent SFRs on average, and nominally the best fit to the full SEDs of the majority of the sample, and we argue that SMC is the most appropriate attenuation for the majority of the sample (except for the highest-mass galaxies). The large observed scatter between \sfrha\ and \sfrsed\ may be plausibly explained by variations in the intrinsic EUV-FUV spectra that are not fully accounted for with standard assumptions.

\end{itemize}

We emphasize that many of the relations between quantities involving continuum and nebular dust corrections depend sensitively on the details of the assumed continuum attenuation curve. In particular, the SMC curve predicts significantly smaller values of reddening and SFR than the ``grayer'' \citet{calzetti2000} curve, and it provides a better fit for many of the galaxies in our sample at $z\sim2.3$. However, a non-negligible fraction is better fit by the \citet{calzetti2000} attenuation curve, and for these galaxies, the application of the SMC curve would lead to severe underestimates of reddening and SFR. Thus, care must be taken when selecting the most appropriate attenuation curve for a given galaxy at high redshift, and in general it is not a good assumption that the same attenuation curve applies to every galaxy at high redshift. 

Perhaps just as importantly, due to large intrinsic variations in $L_{\mathrm{H}\alpha}/L_{\mathrm{UV}}$ modulated by stellar metallicity, IMF, and the importance of binaries, it is unlikely that the same conversion factor between \Ha\ luminosity and SFR applies to every galaxy at high redshift, and we argue that \Ha-based SFRs are highly stochastic, and therefore remain a significant source of systematic errors in estimating SFRs of high-redshift galaxies.

\acknowledgements This work has been supported in part by the NSF through grants AST-0908805 and AST-1313472 (CCS, RLT, ALS), by the JPL President-Director Fund (CCS, RLT), and by an Alfred P. Sloan Research Fellowship (NAR). Finally, the authors wish to recognize and acknowledge the significant cultural role and reverence that the summit of Maunakea has within the indigenous Hawaiian community. We are privileged to have the opportunity to conduct observations from this mountain.

\bibliography{kbss_dust}
\bibliographystyle{aasjournal}

\appendix

\section{SED fitting}\label{sec:sed}
	
\subsection{Fitting Procedure}

The SED fitting uses reddened ``Binary Population and Spectral Synthesis'' \citep[BPASSv2.2;][]{stanway2018} models assuming a constant star formation history (SFH) and a minimum allowed age of 50 Myr. The model SED at each age is redshifted using its measured nebular redshift $z_{\mathrm{neb}}$, reddened from $E(B-V) = 0 - 0.6$ in steps of 0.01 (or $E(B-V) = 0 - 0.3$ in steps of 0.005 if the SMC attenuation curve is assumed), and attenuated blueward of rest-frame 1216 \AA\ by intergalactic medium (IGM) opacity using Monte Carlo modeling. The best-fit normalization of the model at a given age and reddening step is determined by minimizing $\chi^2$ with respect to the observed photometry, and this normalization determines the SFR and stellar mass. The best-fit combination of age and reddening is then taken to be that which minimizes the overall $\chi^2$. This procedure is repeated for each attenuation curve, and we compare the results assuming different curves throughout this work.

The default BPASSv2.2 models do not include the contribution from the nebular continuum emission. As in \citet[][their Figure 3]{steidel2016}, we calculated this contribution relative to the stellar continuum by using the photoionization models that produced the best matches to the observed nebular spectra. This contribution was then added to the BPASSv2.2 models before fitting them to the data. The effect of including the nebular continuum is to make the total continuum slightly redder than the stellar continuum only, thus requiring slightly lower values of \ebmvsed\ to match the data, by 5\% on average.

\subsection{Fit parameters and characteristic uncertainties}\label{sec:sed-fit}
	
\begin{figure*}[tb]
\centering
\plottwo{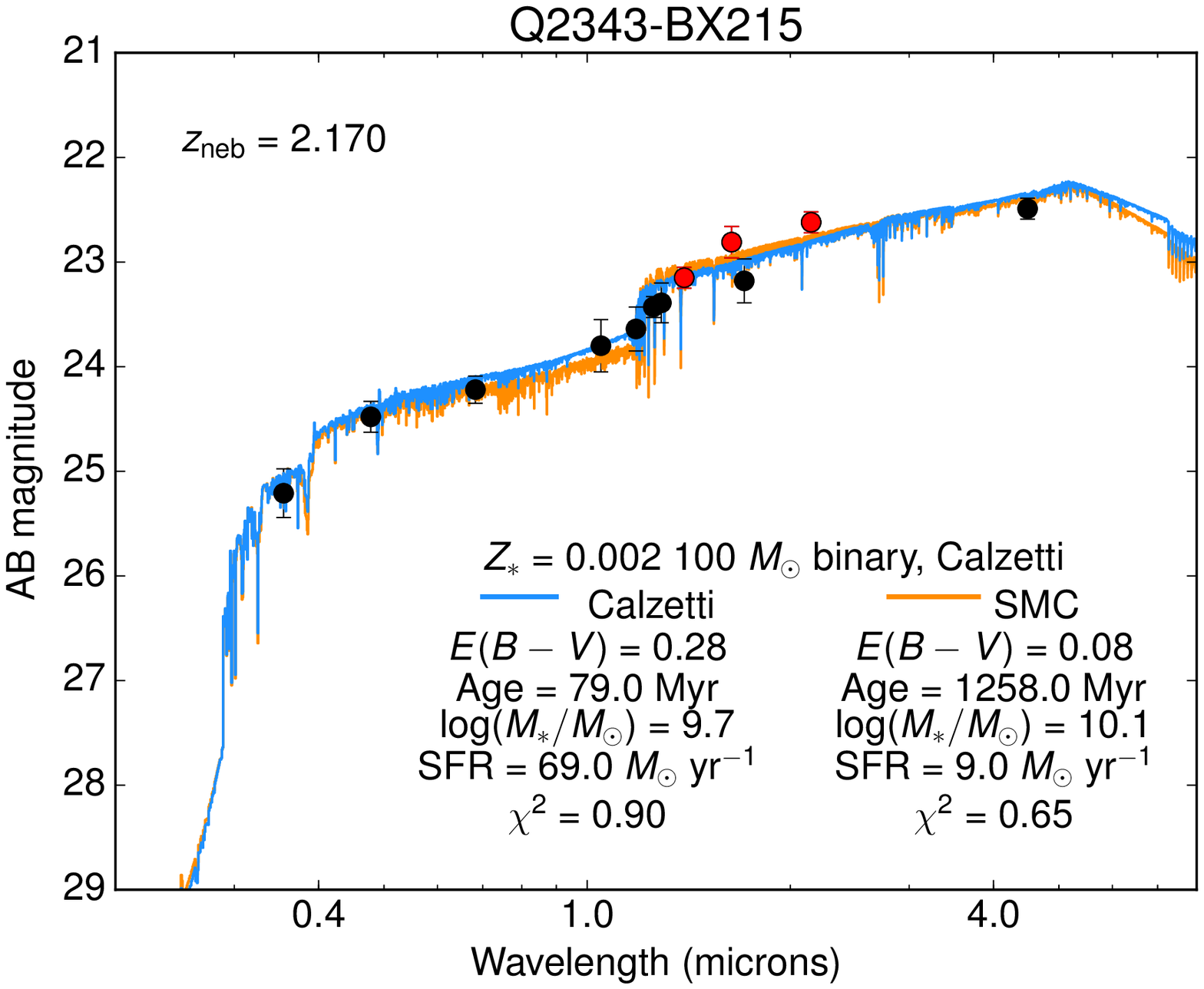}{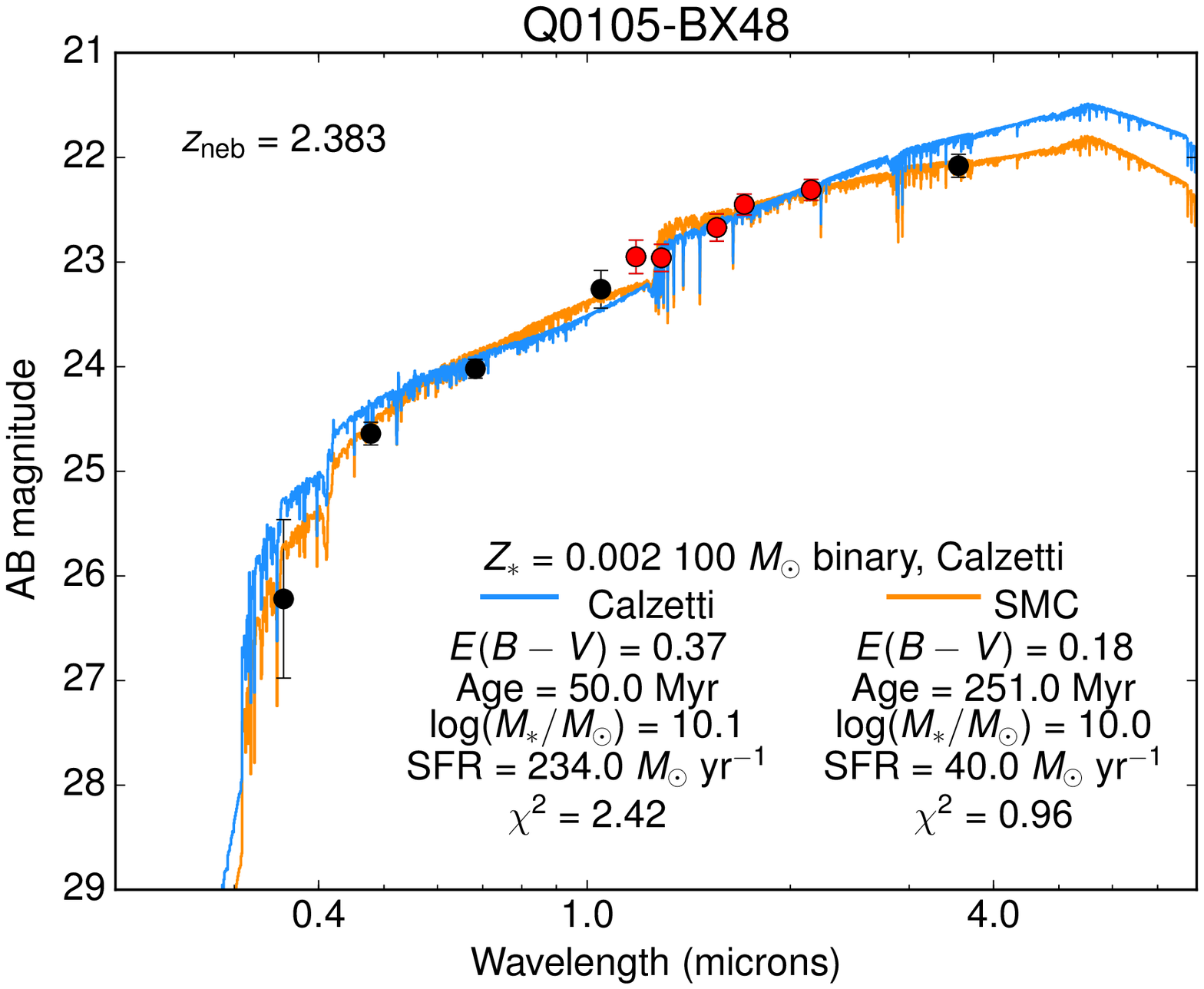}
\caption{Example SEDs for the galaxies Q2343-BX215 (left; $z_{\mathrm{neb}} = 2.170$) and Q0105-BX48 (right; $z_{\mathrm{neb}} = 2.383$). Photometric points are shown in black, and red points are those that have been corrected for the emission line contribution. BPASSv2.2 SEDs are shown for the \citet{calzetti2000} (blue) and SMC (orange) attenuation curves. In the case of Q2343-BX215, both curves give an equally good fit, and while the inferred stellar masses are similar, SMC predicts an older age, lower \ebmvsed, and lower SFR. For Q0105-BX48, however, SMC provides a much better fit. \label{fig:sed}}
\end{figure*}

\begin{figure*}[tb]
\centering
\epsscale{0.5}
\plotone{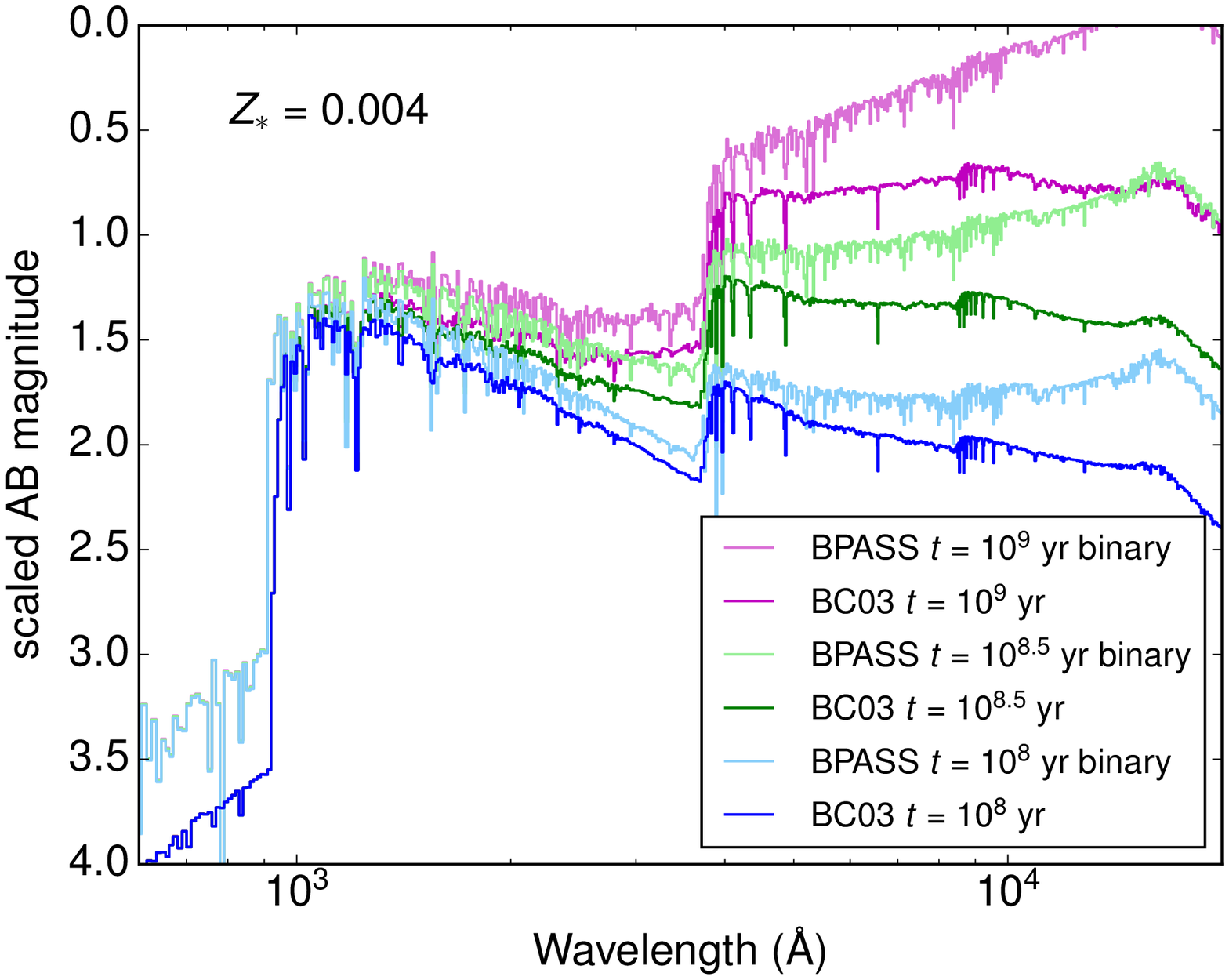}
\caption{Comparison of BPASSv2.2 and \citetalias{bruzual2003} model SEDs, where both models assume a metallicity of $Z_*=0.004$. The SEDs are unreddened and unredshifted, and the BPASSv2.2 models shown here assume an upper mass cutoff of $100M_{\odot}$ with nebular continuum not included. The SEDs are compared at three different ages: $t=10^8$ yr, $t=10^{8.5}$ yr, and $t=10^{9}$ yr. The BPASSv2.2 models are somewhat brighter than \citetalias{bruzual2003} in the far-UV, the region most important for determining \sfrsed. This generally results in lower SFRs given by the BPASSv2.2 models. \label{fig:sed_comp}}
\end{figure*}

\begin{deluxetable}{lDDD}
\tablecaption{Parameters Relative to BPASSv2.2 $Z_* = 0.002$\label{tab:sed}}
\tablehead{\colhead{Quantity} & \multicolumn2c{{BPASSv2.1 $Z_* = 0.002$}} & \multicolumn2c{{BC03 $Z_*=0.004$}} & \multicolumn2c{{BC03 $Z_{\odot}$}}} 
\decimals
\startdata
\ebmvsed\ & $-0.02$ &  $0.00$  & $-0.15$ \\
Age       &  $0.00$ &  $0.00$  & $+0.61$ \\
\mstar\   & $-0.04$ & $+0.10$  & $+0.31$ \\
\sfrsed\  & $-0.04$ & $+0.08$  & $-0.11$ \\
\enddata
\tablecomments{All quoted values are the median difference in dex between the parameters output by the given model and those output by the fiducial model used in this paper, BPASSv2.2 $Z_*=0.002$ $M_{\mathrm{up}} = 100M_{\odot}$. Stellar masses and SFRs have been adjusted to account for the differences in IMF between the BPASS and \citetalias{bruzual2003} models.}
\end{deluxetable}

We estimated a characteristic uncertainty on the SED fit parameters by perturbing each photometric point for each galaxy in the sample 100 times within $1\sigma$, and fitting SEDs to each perturbation using each attenuation curve. We found that the average 68\% confidence intervals for each parameter were: $\pm0.016$ dex for \ebmvsed, $[-0.07, +0.09]$ dex for $\log(t/\mathrm{yr})$, $[-0.07, +0.06]$ dex for $\log(\mathrm{SFR}_{\mathrm{SED}}$, and $\pm0.04$ dex for $\log(M_*)$. We found that 46\% of the galaxies in the sample were equally well fit by both \citet{calzetti2000} and SMC within the uncertainty, which we define as the average 68\% confidence interval on $\chi^2$ for the 100 iterations of the bootstrap.

The lefthand panel of Figure \ref{fig:sed} shows example SED fits for one such galaxy, Q2343-BX215 (left; $z_{\mathrm{neb}} = 2.17$). Both curves gave similar values of $\chi^2$ in this case, and while the inferred stellar masses are similar for both curves, SMC predicts an older age and lower SFR. The righthand panel of Figure \ref{fig:sed} shows example SED fits for the galaxy Q0105-BX48. For this galaxy, SMC gave a significantly lower $\chi^2$.

\subsection{Comparison to other SPS models}

For comparison with our fiducial model, we fit SEDs using the older BPASSv2.1 $Z_*=0.002$ binary models with an upper mass cutoff of $100M_{\odot}$, including nebular continuum. We also fit SEDs using the \citetalias{bruzual2003} $Z_*=0.004$ and $Z_*=0.020$ models (the latter of which is arguably the most commonly used SED model in the literature). We find that the $\chi^2$ of the fits using both of these models were similar to those produced by our fiducial model. A detailed comparison between the four models used for SED fitting is beyond the scope of this work; however, the fractional changes in each fit parameter relative to their BPASSv2.2 $100M_{\odot}$ binary model values are given in Table \ref{tab:sed} (all assuming \citealt{calzetti2000} attenuation). Note that the \citetalias{bruzual2003} models do not include the contribution from the nebular continuum.

We also fit SEDs using the BPASS models with an upper mass cutoff of $300M_{\odot}$, and found that changing the upper mass cutoff has almost no effect on any of the inferred parameters due to the relative insensitivity of the integrated SED in the range $0.35-4.5$\micron\ (or $\sim$1200\AA$-1.5$\micron\ in the rest frame) to the most massive stars; the primary difference between the two models is in the (unobservable) EUV spectrum, and thus, the predicted strengths of nebular emission lines and line ratios. We selected the $100M_{\odot}$ model as our fiducual model because the \citet{calzetti2000} curve assumes an upper mass IMF cutoff of $100M_{\odot}$. Table \ref{tab:sed} compares the inferred SED fit parameters for the three models discussed in this paper.

Figure \ref{fig:sed_comp} compares the model SEDs for BPASSv2.2 and \citetalias{bruzual2003} for three choices of stellar population age, where both models assume a metallicity of $Z_*=0.004$, and nebular continuum is not included for the BPASSv2.2 models. The BPASSv2.2 models are somewhat brighter than \citetalias{bruzual2003} blueward of the Balmer break at $t=10^8$ yr, which is close to the median age predicted by the \citetalias{bruzual2003} $Z_*=0.020$ models.

\section{Spectral synthesis fitting to rest-UV spectra}\label{sec:fitting}

\begin{figure*}[tbh]
\centering
\gridline{\fig{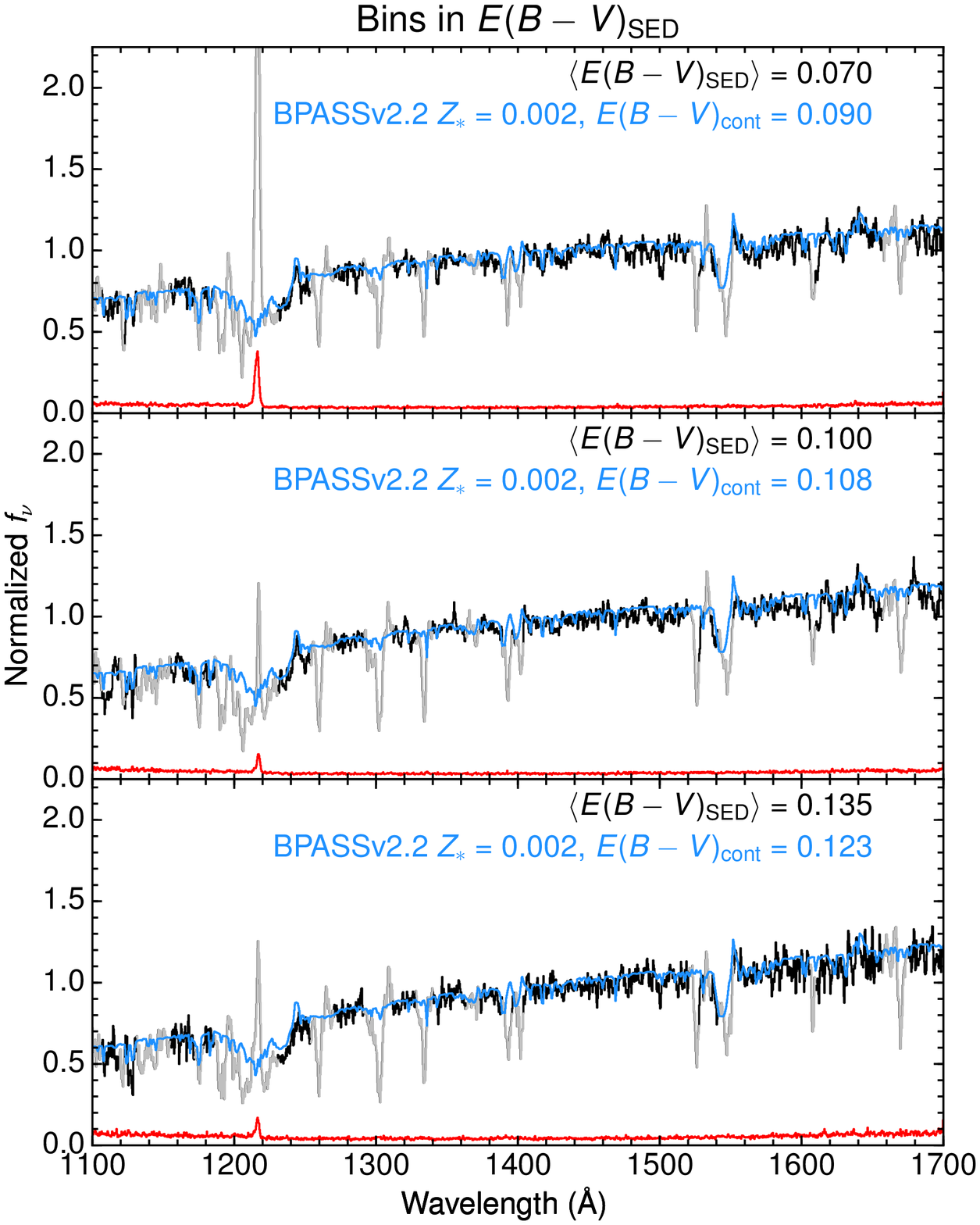}{0.45\textwidth}{(a)}
          \fig{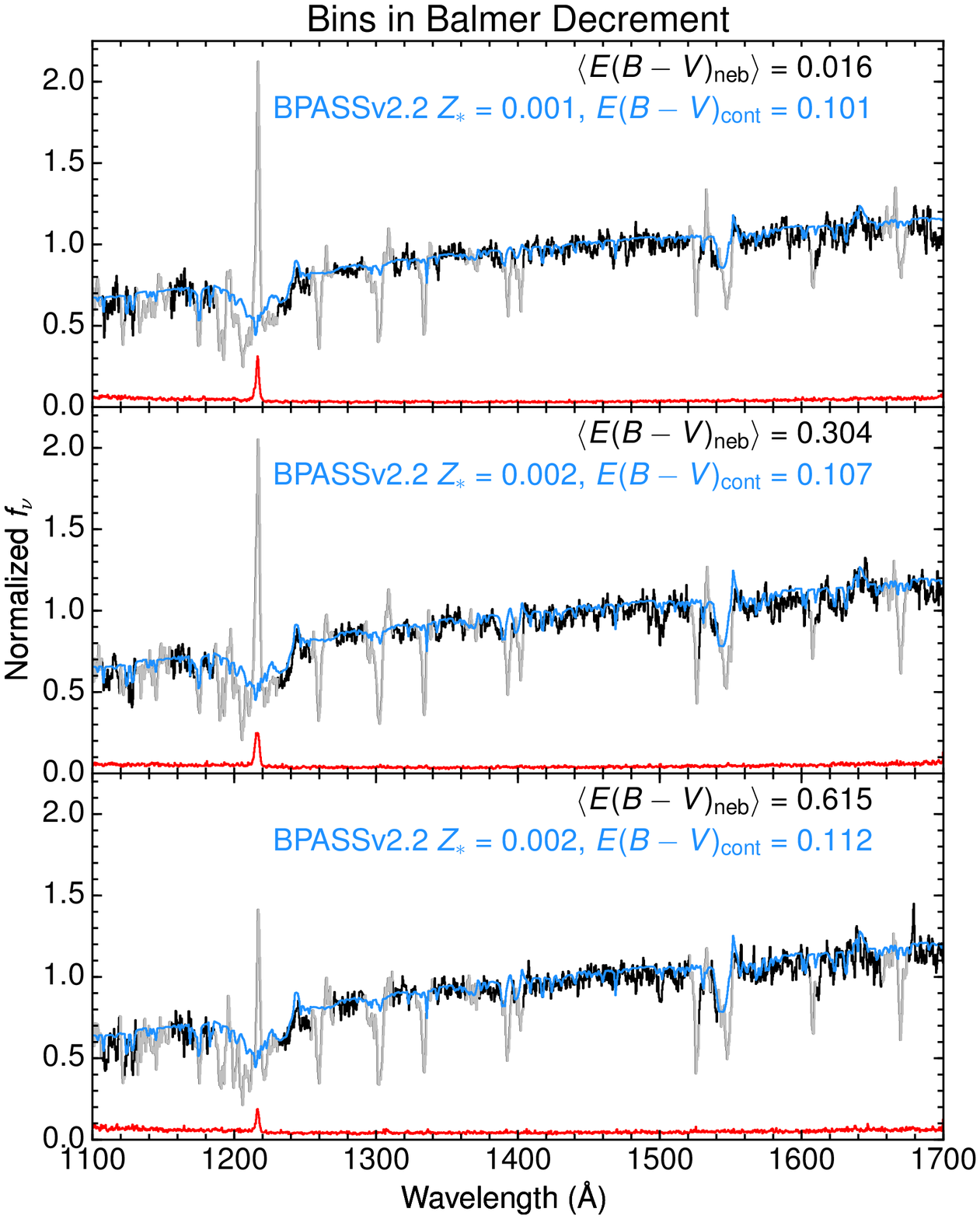}{0.45\textwidth}{(b)}}
\gridline{\fig{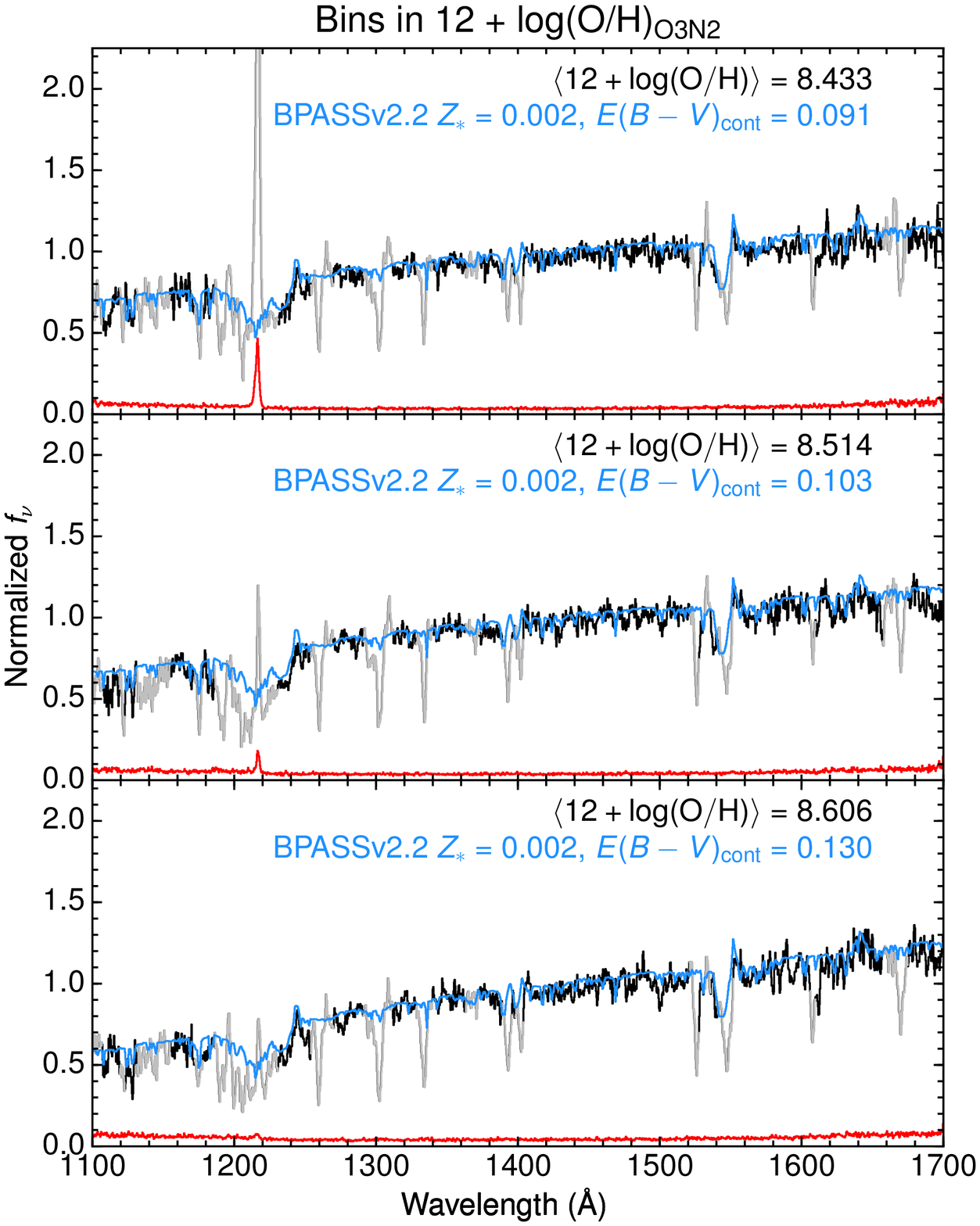}{0.45\textwidth}{(c)}
          \fig{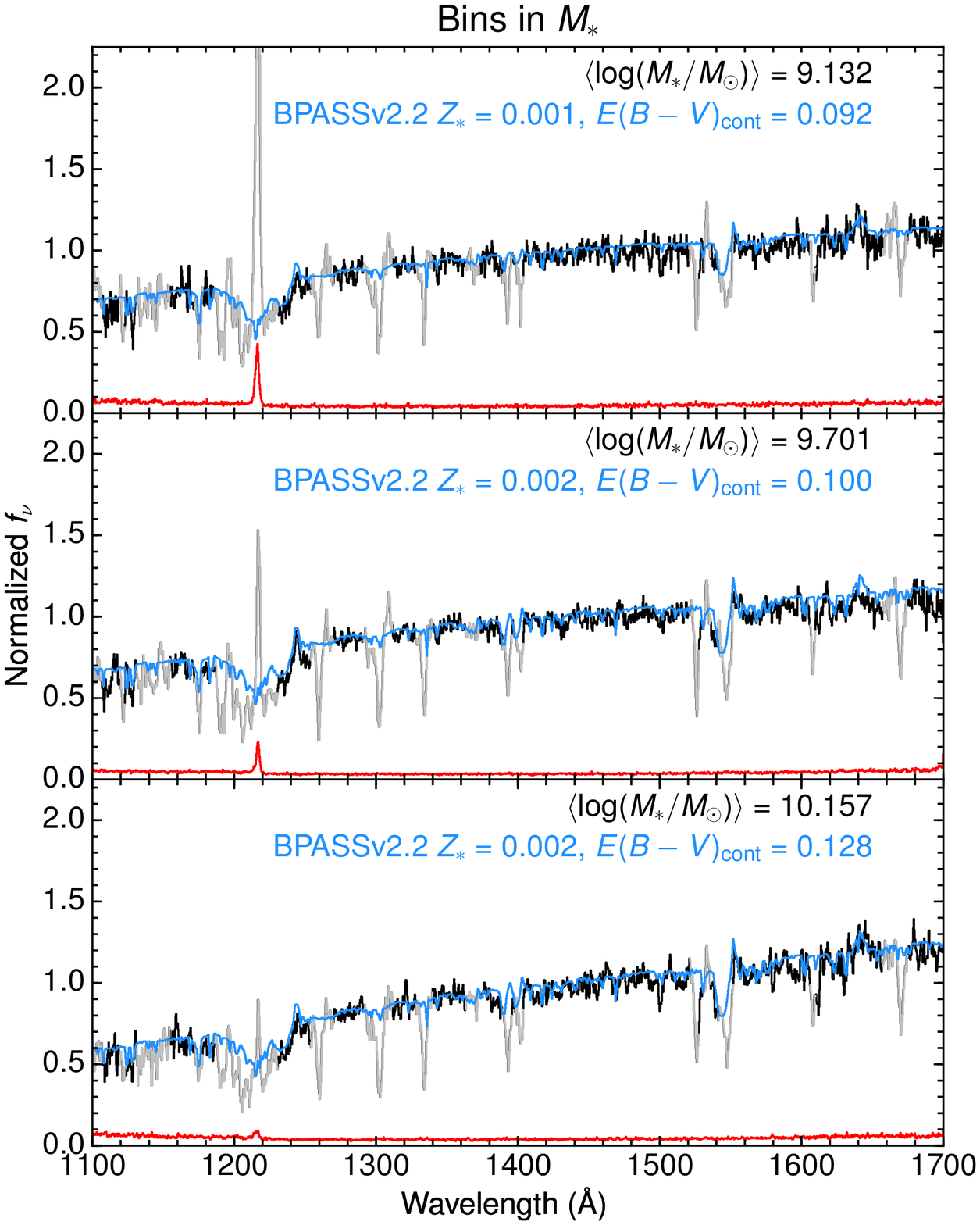}{0.45\textwidth}{(d)}}
	\caption{Stacked composite rest-frame UV spectra of the LRIS+MOSFIRE sample, in bins of a) \ebmvsed, b) BD (converted to \ebmvneb), c) \Zneb, d) $M_*$, e) \sfrha, and f) \sfrsed. Bootstrapped error spectra are shown in red. Prior to stacking, the spectra were normalized by the median flux in the range $1400-1500$ \AA. Superposed are the best-fit population synthesis models from BPASSv2.2 (blue), reddened by the best-fit \ebmvcont\ (assuming an SMC attenuation curve), which was calculated by minimizing $\chi^2$ with respect to the observed spectrum, after masking regions containing nebular emission lines or interstellar absorption lines (gray). \label{fig:uvfits}}
	\end{figure*}

\begin{figure*}[tbh]\ContinuedFloat
    \centering
    \gridline{\fig{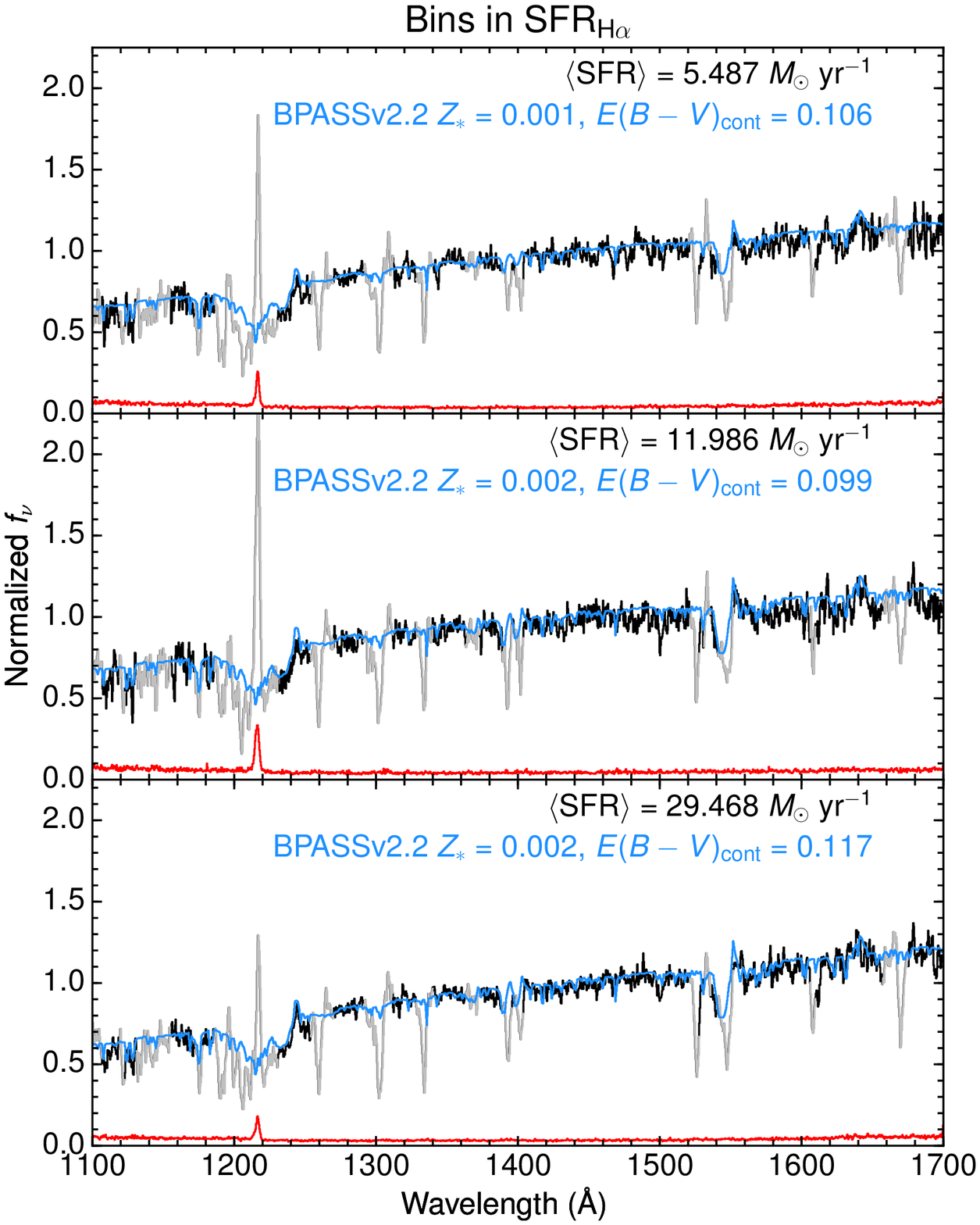}{0.45\textwidth}{(e)}
              \fig{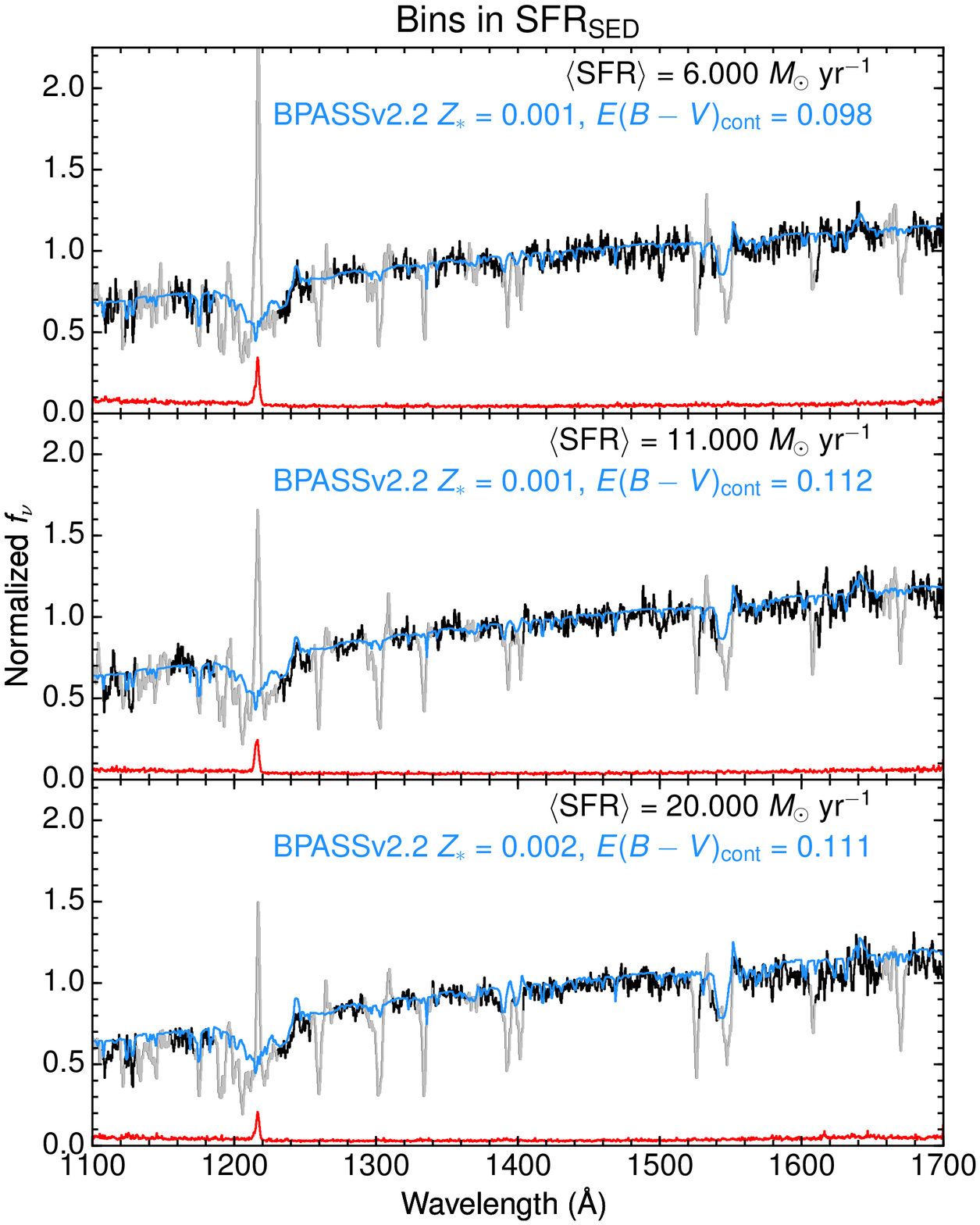}{0.45\textwidth}{(f)}}
    \caption{(Continued)}
    \end{figure*}
    
\subsection{Generating the composite rest-UV spectra}\label{sec:stacks}

To generate the composites described in Section \ref{sec:rest-uv}, the UV spectra were shifted to the rest frame using $z_{\mathrm{neb}}$, and the sample was sorted into three equal-number bins of six quantities: \ebmvsed, \ebmvneb, \Zneb, $M_*$, \sfrha, and \sfrsed. The spectra were then spline-interpolated onto a rest wavelength scale of 0.35 \AA/pixel, and two positive and negative extrema (four points total) were rejected at each dispersion pixel. The composite spectra were generated by normalizing each spectrum by the median flux density in the range $1400-1500$ \AA\ and averaging without weighting. The error spectra were generated by a bootstrap technique similar to that described by \citet{trainor2015}; over 1000 iterations, random spectra were drawn with replacement (the same number of spectra as in the bin) and stacked. Outlier rejection was performed on the bootstrap spectra in the same manner as the data. The standard deviation of the values at each pixel over all 1000 iterations was used to define the error spectrum associated with each stack. Thus, the error spectrum encompasses both the observational uncertainties in individual spectra and the intrinsic variation among galaxies in a given stack.

The composites were corrected for the mean opacity of the IGM due to neutral hydrogen along the line of sight using measurements from \citet{rudie2013}, for a source with $z = 2.40$ (this is close to the mean redshift of the sample; see Table \ref{tab:sample}). Details of this correction are described by \citet{steidel2016}, but we note that the IGM correction only affects the spectrum shortward of 1216 \AA.

\subsection{Spectral synthesis model details}\label{sec:models}

For fitting to the observed data, we used the far-UV spectra generated by the BPASSv2.2 model suite \citep{stanway2018} assuming a constant star formation history with an age of $10^8$ yr (see \citealt{steidel2016} for details). We considered only models with an upper mass cutoff of $100M_{\odot}$ which include binary evolution, with stellar metallicities $Z_*=(0.001, 0.002, 0.003, 0.004, 0.006, 0.008, 0.010, 0.014)$. The stellar metallicity of the model is allowed to vary independently of any knowledge of the gas-phase O/H in the three bins as a means of accounting for non-solar abundance ratios of O relative to Fe, where $Z_*$ maps to Fe/H in the stars \citep{steidel2016,strom2018}. As with the SED fitting (Appendix \ref{sec:sed}), we self-consistently included the nebular continuum contribution. The model spectra were normalized by the median flux in the range $1400-1500$ \AA\ in order to match the data. 

For comparison with the observed spectra, the model spectra were spline-interpolated onto the pixel scale of each composite. We varied only the amount of stellar continuum reddening, which determines the ``slope'' of the spectrum. The model spectra were reddened with the \citet{calzetti2000} and SMC attenuation curves, and the continuum color excess \ebmvcont\ was varied over the range $0.000 \leq E(B-V)_{\mathrm{cont}} \leq 0.600$ in steps of $0.001$. Spectral regions corresponding to nebular emission lines and interstellar absorption lines were excluded from fitting (see \citealt{steidel2016} for details). The best-fit combination of $Z_*$ and \ebmvcont\ for each spectral synthesis model was determined by computing the total $\chi^2$, summed over all unmasked pixels. The minimum $\chi^2$ was then compared for each attenuation curve, for each of the six binned quantities.

Figure \ref{fig:uvfits} shows stacked composite rest-frame UV spectra of the LRIS+MOSFIRE sample, in bins of a) \ebmvsed, b) BD [converted to \ebmvneb], c) \Zneb, d) $M_*$, e) \sfrha, and f) \sfrsed, overlaid with the best-fit spectral synthesis models from BPASSv2.2 (blue). For reasons of space, only the fits using the SMC curve are shown. The strong relationship between Lyman-$\alpha$ line strengths and various properties apparent in Figure \ref{fig:uvfits} will be discussed in a future paper.

\end{document}